\documentstyle[12pt,epsfig]{article}

\newcommand{\agt}{\,\rlap{\lower 3.5 pt \hbox{$\mathchar \sim$}} \raise 1pt
 \hbox {$>$}\,}
\newcommand{\alt}{\,\rlap{\lower 3.5 pt \hbox{$\mathchar \sim$}} \raise 1pt
 \hbox {$<$}\,}
\newcommand{\Arcosh}{\mathop{\mbox{Arcosh}}\nolimits}

\catcode`@=11
\newcount\@tempcntc
\def\@citex[#1]#2{\if@filesw\immediate\write\@auxout{\string\citation{#2}}\fi
  \@tempcnta\z@\@tempcntb\m@ne\def\@citea{}\@cite{\@for\@citeb:=#2\do
    {\@ifundefined
       {b@\@citeb}{\@citeo\@tempcntb\m@ne\@citea\def\@citea{,}{\bf ?}\@warning
       {Citation `\@citeb' on page \thepage \space undefined}}%
    {\setbox\z@\hbox{\global\@tempcntc0\csname b@\@citeb\endcsname\relax}%
     \ifnum\@tempcntc=\z@ \@citeo\@tempcntb\m@ne
       \@citea\def\@citea{,}\hbox{\csname b@\@citeb\endcsname}%
     \else
      \advance\@tempcntb\@ne
      \ifnum\@tempcntb=\@tempcntc
      \else\advance\@tempcntb\m@ne\@citeo
      \@tempcnta\@tempcntc\@tempcntb\@tempcntc\fi\fi}}\@citeo}{#1}}
\def\@citeo{\ifnum\@tempcnta>\@tempcntb\else\@citea\def\@citea{,}%
  \ifnum\@tempcnta=\@tempcntb\the\@tempcnta\else
   {\advance\@tempcnta\@ne\ifnum\@tempcnta=\@tempcntb \else \def\@citea{--}\fi
    \advance\@tempcnta\m@ne\the\@tempcnta\@citea\the\@tempcntb}\fi\fi}
\catcode`@=12
\begin{document}
\title{\vskip-3cm{\baselineskip14pt
\centerline{\normalsize CERN-TH/2002-167\hfill ISSN 0418-9833}
\centerline{\normalsize DESY 02-101\hfill}
\centerline{\normalsize hep-ph/0208104\hfill}
\centerline{\normalsize August 2002\hfill}}
\vskip1.5cm
Associated Production of Heavy Quarkonia and Electroweak Bosons at Present and
Future Colliders}
\author{{\sc Bernd A. Kniehl}\thanks{Permanent address: II. Institut f\"ur
Theoretische Physik, Universit\"at Hamburg, Luruper Chaussee 149, 22761
Hamburg, Germany.}\\
{\normalsize CERN, Theoretical Physics Division, 1211 Geneva 23,
Switzerland}\\
\\
{\sc Caesar P. Palisoc}\\
{\normalsize National Institute of Physics, University of the Philippines,}\\
{\normalsize Diliman, Quezon City 1101, Philippines}\\
\\
{\sc Lennart Zwirner}\\
{\normalsize II. Institut f\"ur Theoretische Physik, Universit\"at Hamburg,}\\
{\normalsize Luruper Chaussee 149, 22761 Hamburg, Germany}}

\date{}

\maketitle

\thispagestyle{empty}

\begin{abstract}
We investigate the associated production of heavy quarkonia, with
angular-momentum quantum numbers
${}^{2S+1}L_J={}^1\!S_0,{}^3\!S_1,{}^1\!P_1,{}^3\!P_J$ ($J=0,1,2$), and
photons, $Z$ bosons, and $W$ bosons in photon-photon, photon-hadron, and
hadron-hadron collisions within the factorization formalism of nonrelativistic
quantum chromodynamics providing all contributing partonic cross sections in
analytic form.
In the case of photoproduction, we also include the resolved-photon
contributions.
We present numerical results for the processes involving $J/\psi$ and
$\chi_{cJ}$ mesons appropriate for the Fermilab Tevatron, CERN LHC, DESY TESLA,
operated in the $e^+e^-$ and $\gamma\gamma$ modes, and DESY THERA.

\medskip

\end{abstract}

\newpage

\section{Introduction}
\label{sec:one}

Since its discovery in 1974, the $J/\psi$ meson has provided a useful
laboratory for quantitative tests of quantum chromodynamics (QCD) and, in
particular, of the interplay of perturbative and nonperturbative phenomena.
The factorization formalism of nonrelativistic QCD (NRQCD) \cite{bbl} provides
a rigorous theoretical framework for the description of heavy-quarkonium
production and decay.
This formalism implies a separation of short-distance coefficients, which can 
be calculated perturbatively as expansions in the strong-coupling constant
$\alpha_s$, from long-distance matrix elements (MEs), which must be extracted
from experiment.
The relative importance of the latter can be estimated by means of velocity
scaling rules, {\it i.e.}, the MEs are predicted to scale with a definite
power of the heavy-quark ($Q$) velocity $v$ in the limit $v\ll1$.
In this way, the theoretical predictions are organized as double expansions in
$\alpha_s$ and $v$.
A crucial feature of this formalism is that it takes into account the complete
structure of the $Q\overline{Q}$ Fock space, which is spanned by the states
$n={}^{2S+1}L_J^{(c)}$ with definite spin $S$, orbital angular momentum $L$,
total angular momentum $J$, and color multiplicity $c=1,8$.
In particular, this formalism predicts the existence of color-octet (CO)
processes in nature.
This means that $Q\overline{Q}$ pairs are produced at short distances in
CO states and subsequently evolve into physical, color-singlet (CS) quarkonia
by the nonperturbative emission of soft gluons.
In the limit $v\to0$, the traditional CS model (CSM) \cite{ber} is recovered.
The greatest triumph of this formalism was that it was able to correctly 
describe \cite{bra} the cross section of inclusive charmonium
hadroproduction measured in $p\overline{p}$ collisions at the Fermilab
Tevatron \cite{abe}, which had turned out to be more than one order of
magnitude in excess of the theoretical prediction based on the CSM.

In order to convincingly establish the phenomenological significance of the
CO processes, it is indispensable to identify them in other kinds of
high-energy experiments as well.
Studies of charmonium production in $ep$ photoproduction, $ep$ and $\nu N$
deep-inelastic scattering, $e^+e^-$ annihilation, $\gamma\gamma$ collisions,
and $b$-hadron decays may be found in the literature; see Ref.~\cite{fle} and
references cited therein.
Furthermore, the polarization of charmonium, which also provides a sensitive
probe of CO processes, was investigated \cite{ben,bkl}.
Until very recently, none of these studies was able to prove or disprove the
NRQCD factorization hypothesis \cite{bbl}.
However, preliminary data of $\gamma\gamma\to J/\psi+X$ taken by the DELPHI
Collaboration \cite{delphi} at LEP2 provide first independent evidence for it
\cite{gg}.

In most NRQCD studies of charmonium production in high-energy particle
collisions, the charmonium state is produced either singly or in association
with a hadron jet ($j$), which originates from an outgoing quark ($q$) or
gluon ($g$), because this is bound to yield the largest cross sections
\cite{ber,bra,ben,bkl,gg}.
However, also the associated production of $J/\psi$ mesons and prompt photons
($\gamma$) was considered, in photon-photon \cite{ma,jap,kla,god},
photon-hadron \cite{meh,cac}, and hadron-hadron collisions \cite{cac,kim}.
Although the cross sections of the latter processes are suppressed by a factor
of $\alpha/\alpha_s$, where $\alpha$ is Sommerfeld's fine-structure constant,
relative to the case of $J/\psi+j$ associated production, these processes
exhibit some attractive features.
On the one hand, $J/\psi+\gamma$ associated production in photon-photon
collisions dominantly proceeds through the CS partonic subprocesses of direct
and doubly-resolved photoproduction,
$\gamma\gamma\to c\overline{c}\left[{}^3\!S_1^{(1)}\right]\gamma$ and
$gg\to c\overline{c}\left[{}^3\!S_1^{(1)}\right]\gamma$, respectively, and
thus allows for an independent determination of the CS ME
$\langle{\cal O}^{J/\psi}[{}^3\!S_1^{(1)}]\rangle$, which is usually extracted
from the measured leptonic annihilation rate of the $J/\psi$ meson.
On the other hand, $J/\psi+\gamma$ associated production in photon-hadron
collisions provides a good handle on the CO mechanism because the partonic
subprocesses of direct photoproduction, $\gamma g\to c\overline{c}[n]\gamma$
with $n={}^1\!S_0^{(8)},{}^3\!S_1^{(8)},{}^3\!P_J^{(8)}$, are pure CO
processes \cite{meh,cac}.
Finally, for small $J/\psi$ transverse momentum, $J/\psi+\gamma$ associated
hadroproduction chiefly proceeds through the CS partonic subprocess
$gg\to c\overline{c}\left[{}^3\!S_1^{(1)}\right]\gamma$ and thus lends
itself as a clean probe of the gluon density inside the proton \cite{cac,kim}.

The purpose of this paper is to study the associated production of heavy
quarkonia with the intermediate bosons, $W$ and $Z$, in photon-photon,
photon-hadron, and hadron-hadron collisions to leading order (LO) in the NRQCD
factorization formalism.
We consider all experimentally established heavy quarkonia, which are
classified by their angular-momentum quantum numbers
${}^{2S+1}L_J={}^1\!S_0,{}^3\!S_1,{}^1\!P_1,{}^3\!P_J$ with $J=0,1,2$.
In the case of charmonium, these correspond to the $\eta_c$, $J/\psi$
($\psi^\prime$, $\psi(3S)$, {\it etc.}), $h_c$, and $\chi_{cJ}$ mesons,
respectively.
We present all contributing partonic cross sections in analytic form.
As a by-product of our analysis, we recover the corresponding formulas for the
associated production of heavy quarkonia with prompt photons, by appropriately
adjusting the couplings and mass of the $Z$ boson.
In our numerical study, we concentrate on those charmonia which can be most
straightforwardly identified experimentally, namely the $J/\psi$ and
$\chi_{cJ}$ mesons, through their leptonic and radiative decays, respectively.
Specifically, we assess the feasibilities of the $p\overline{p}$ synchrotron
Tevatron, which is in operation at Fermilab, the $pp$ synchrotron LHC, which
is under construction at CERN, the $e^+e^-$ linear collider (LC) TESLA, which
is being developed and planned at DESY and can be operated in the $e^+e^-$
\cite{tesla} and $\gamma\gamma$ \cite{compton} modes, and the $pe^\pm$
collider THERA, which uses the TESLA lepton beam and the HERA proton beam
\cite{thera}, to produce $J/\psi$ and $\chi_{cJ}$ mesons in association with
prompt photons, $W$ bosons, and $Z$ bosons.
These colliders all operate with high luminosities at high energies and, at
first sight, have the potential to produce these final states with reasonable
rates.
Furthermore, if one selects the leptonic decays of the $J/\psi$ meson and
intermediate bosons, then these final states lead to spectacular signals,
consisting of energetic isolated prompt photons, charged leptons, and possibly
missing four-momentum, which should be easy to detect even at hadron
colliders.
A first step in this direction was recently undertaken in Ref.~\cite{lee},
which contains a numerical study of $\Upsilon+W$ and $\Upsilon+Z$
hadroproduction at the Tevatron and the LHC.

For completeness, we also provide numerical results for the cross sections of
the production of $J/\psi$ and $\chi_{cJ}$ mesons together with standard-model
(SM) Higgs bosons, which are suppressed by the smallness of the charm Yukawa
coupling.
In fact, explicit analysis reveals that these cross sections are too small to
produce any signals for the envisaged luminosities of the considered
colliders, even in the case of the LHC.
By the same token, the observation of events where a scalar boson is produced
together with charmonium would signal new physics beyond the SM.

This paper is organized as follows.
In Sec.~\ref{sec:two}, we present our analytic results and explain how to
evaluate the cross sections of the various processes enumerated above.
The contributing partonic cross sections are collected in the Appendix.
In Sec.~\ref{sec:three}, we present our numerical results.
Our conclusions are summarized in Sec.~\ref{sec:four}.

\section{Analytic results}
\label{sec:two}

In this section, we explain how to calculate the cross sections of the
associated production of a heavy quarkonium state $C$ and an electroweak boson
$D$ by photoproduction in $e^+e^-$ and $pe^\pm$ collisions and by
hadroproduction in $p\overline{p}$ and $pp$ collisions.

To start with, we consider $e^+e^-$ collisions, where photons are unavoidably
generated by hard initial-state bremsstrahlung.
At a high-energy $e^+e^-$ LC, an additional source of hard photons is provided
by beamstrahlung, the synchrotron radiation emitted by one of the colliding
bunches in the field of the opposite bunch.
Bremsstrahlung and beamstrahlung both occur in the $e^+e^-$ mode, and their
energy spectra and must be added coherently.
The highest possible photon energies with large enough luminosity may be
achieved by converting the $e^+e^-$ LC into a $\gamma\gamma$ collider via
back-scattering of high-energetic laser light off the electron and positron
beams.

The photons can interact either directly with the quarks participating in the
hard-scattering process (direct photoproduction) or via their quark and gluon
content (resolved photoproduction).
Thus, the inclusive process $e^+e^-\to e^+e^-CD+X$, where $X$ denotes the
hadronic remnant, receives contributions from the direct, singly-resolved, and
doubly-resolved channels.
All three contributions are formally of the same order in the perturbative
expansion and must be included.
This may be understood by observing that the parton density functions (PDFs)
of the photon have a leading behavior proportional to
$\alpha\ln(M^2/\Lambda_{\rm QCD}^2)\propto\alpha/\alpha_s$, where $M$ is the
factorization scale and $\Lambda_{\rm QCD}$ is the asymptotic scale parameter
of QCD.

Let us now describe how to calculate the cross section.
We take the electron and positron to be massless and denote the masses of $C$
and $D$ by $m_C$ and $m_D$, respectively.
Let $\sqrt S$ be the energy of the initial state, $y_C$ and $y_D$ the
rapidities of $C$ and $D$, respectively, and $p_T$ their common transverse
momentum in the center-of-mass (c.m.) frame of the collision.
Invoking the Weizs\"acker-Williams approximation (WWA) for electromagnetic
bremsstrahlung \cite{wei}, its analogues for beamstrahlung \cite{che,bar} and
Compton back-scattering \cite{gin,tel}, and the factorization theorems of the
QCD parton model \cite{dew} and NRQCD \cite{bbl}, the differential cross
section of $e^+e^-\to e^+e^-CD+X$ can be written as
\begin{eqnarray}
\lefteqn{\frac{d^3\sigma}{dp_T^2\,dy_C\,dy_D}(e^+e^-\to e^+e^-CD+X)
=\int_{\overline{x}_+}^1dx_+\,f_{\gamma/e}(x_+)
\int_{\overline{x}_-}^1dx_-\,f_{\gamma/e}(x_-)}
\nonumber\\
&&{}\times
\sum_{a,b,n}x_af_{a/\gamma}(x_a,M)x_bf_{b/\gamma}(x_b,M)
\langle{\cal O}^C[n]\rangle\frac{d\sigma}{dt}(ab\to Q\overline{Q}[n]+D),
\label{eq:master}
\end{eqnarray}
where it is summed over the active partons $a,b=\gamma,g,q,\overline{q}$,
$f_{\gamma/e}(x_\pm)$ are the photon flux functions,
$f_{a/\gamma}(x_a,M)$ and $f_{b/\gamma}(x_b,M)$ are the PDFs of the photon,
$\langle{\cal O}^C[n]\rangle$ are the MEs of $C$,
$(d\sigma/dt)(ab\to Q\overline{Q}[n]+D)$ are the differential partonic cross 
sections, and $x_\pm$ and $x_{a,b}$ are the fractions of longitudinal momentum
that the emitted particles receive from the emitting ones.
With the definition $f_{\gamma/\gamma}(x_\gamma,M)=\delta(1-x_\gamma)$, the
direct, singly-resolved, and doubly-resolved channels are all accommodated.
We have
\begin{eqnarray}
\overline{x}_\pm&=&\frac{m_T^C\exp(\pm y_C)+m_T^D\exp(\pm y_D)}{\sqrt S},
\label{eq:xpm}\\
x_{a,b}&=&
\frac{m_T^C\exp(\pm y_C)+m_T^D\exp(\pm y_D)}{x_\pm\sqrt S},
\label{eq:xab}
\end{eqnarray}
where $m_T^C=\sqrt{m_C^2+p_T^2}$ is the transverse mass of $C$ and similarly
for $D$.

The partonic Mandelstam variables $s=(p_a+p_b)^2$, $t=(p_a-p_C)^2$, and
$u=(p_a-p_D)^2$ can be expressed in terms of $p_T$, $y_C$, and $y_D$, as
\begin{eqnarray}
s&=&\left(m_T^C\right)^2+\left(m_T^D\right)^2+2m_T^Cm_T^D\cosh(y_C-y_D),
\nonumber\\
t&=&-p_T^2-m_T^Cm_T^D\exp(y_D-y_C),
\nonumber\\
u&=&-p_T^2-m_T^Cm_T^D\exp(y_C-y_D),
\end{eqnarray}
respectively.
Notice that $s+t+u=m_C^2+m_D^2$ and $sp_T^2=tu-m_C^2m_D^2$.
The kinematically allowed ranges of $S$, $p_T$, $y_C$, and $y_D$ are
\begin{eqnarray}
S&\ge&(m_C+m_D)^2,
\nonumber\\
0&\le&p_T\le\frac{1}{2}\sqrt{\frac{\lambda\left(S,m_C^2,m_D^2\right)}{S}},
\nonumber\\
|y_C|&\le&\Arcosh\frac{S+m_C^2-m_D^2}{2\sqrt Sm_T^C},
\nonumber\\
-\ln\frac{\sqrt S-m_T^C\exp(-y_C)}{m_T^D}&\le&y_D\le
\ln\frac{\sqrt S-m_T^C\exp(y_C)}{m_T^D},
\end{eqnarray}
where $\lambda(x,y,z)=x^2+y^2+z^2-2(xy+yz+zx)$ is K\"all\'en's function.

Sometimes it may be interesting to consider the cross section differential in
the $CD$ invariant mass $m_{CD}=\sqrt s$ rather than in $p_T$.
Using
\begin{equation}
p_T^2=\frac{r\cosh(y_C-y_D)-m_{CD}^2}{2\sinh^2(y_C-y_D)},
\label{eq:pt}
\end{equation}
where $r=\sqrt{m_{CD}^4+\left(m_C^2-m_D^2\right)^2\sinh^2(y_C-y_D)}$, we find
\begin{equation}
\frac{dp_T^2}{dm_{CD}^2}=\frac{m_{CD}^2\cosh(y_C-y_D)-r}{2r\sinh^2(y_C-y_D)}.
\label{eq:mhb}
\end{equation}
Equations~(\ref{eq:pt}) and (\ref{eq:mhb}) are regular for $y_C=y_D$ and read
then
\begin{eqnarray}
p_T^2&=&\frac{\lambda\left(S,m_C^2,m_D^2\right)}{4S},
\nonumber\\
\frac{dp_T^2}{dm_{CD}^2}&=&
\frac{1}{4}\left[1-\left(\frac{m_C^2-m_D^2}{m_{CD}^2}\right)^2\right],
\end{eqnarray}
respectively.
Multiplying Eq.~(\ref{eq:master}) with Eq.~(\ref{eq:mhb}), we obtain
$\left(d^3\sigma/dm_{CD}^2\,dy_C\,dy_D\right)(e^+e^-\to e^+e^-CD+X)$.
The kinematically allowed range of $m_{CD}$ is $m_C+m_D\le m_{CD}\le\sqrt S$.

We work in the fixed-flavor-number scheme, {\it i.e.}, we have $n_f=3$ active
quark flavors $q=u,d,s$ if $Q=c$ and $n_f=4$ active quark flavors
$q=u,d,s,c$ if $Q=b$.
As required by parton-model kinematics, we treat the $q$ quarks as massless.
To LO in $v$, we need to include the $c\overline{c}$ Fock states
$n={}^1\!S_0^{(1)},{}^1\!S_0^{(8)},{}^3\!S_1^{(8)},{}^1\!P_1^{(8)}$ if
$H=\eta_c$;
$n={}^3\!S_1^{(1)},{}^1\!S_0^{(8)},{}^3\!S_1^{(8)},{}^3\!P_J^{(8)}$ if
$H=J/\psi,\psi^\prime,\psi(3S),\ldots$;
$n={}^1\!P_1^{(1)},{}^1\!S_0^{(8)}$ if $H=h_c$; and
$n={}^3\!P_J^{(1)},{}^3\!S_1^{(8)}$ if $H=\chi_{cJ}$, where $J=0,1,2$
\cite{bbl}.
Their MEs satisfy the multiplicity relations
\begin{eqnarray}
\left\langle{\cal O}^{\psi(nS)}\left[{}^3\!P_J^{(8)}\right]\right\rangle
&=&(2J+1)
\left\langle{\cal O}^{\psi(nS)}\left[{}^3\!P_0^{(8)}\right]\right\rangle,
\nonumber\\
\left\langle{\cal O}^{\chi_{cJ}}\left[{}^3\!P_J^{(1)}\right]\right\rangle
&=&(2J+1)
\left\langle{\cal O}^{\chi_{c0}}\left[{}^3\!P_0^{(1)}\right]\right\rangle,
\nonumber\\
\left\langle{\cal O}^{\chi_{cJ}}\left[{}^3\!S_1^{(8)}\right]\right\rangle
&=&(2J+1)
\left\langle{\cal O}^{\chi_{c0}}\left[{}^3\!S_1^{(8)}\right]\right\rangle,
\label{eq:mul}
\end{eqnarray}
which follow to LO in $v$ from heavy-quark spin symmetry.
The assignments for the various bottomonia are analogous.

The cross section of $pe^\pm\to e^\pm CD+X$ in photoproduction emerges from
Eq.~(\ref{eq:master}) by substituting $f_{\gamma/e}(x_+)=\delta(1-x_+)$ and
replacing the photon PDFs $f_{a/\gamma}(x_a,M)$ with their proton counterparts
$f_{a/p}(x_a,M)$.
It receives contributions from the direct and resolved channels.
Here, it is understood that $f_{\gamma/e}(x_-)$ only accounts for
electromagnetic bremsstrahlung.
To conform with HERA conventions, we take the rapidities to be positive in the
proton flight direction.
The rapidity $y_C^{\rm lab}$ of $C$ in the laboratory frame, where the proton
and electron have energies $E_p$ and $E_e$, respectively, is related to $y_C$
by
\begin{equation}
y_C^{\rm lab}=y_C+\frac{1}{2}\ln\frac{E_p}{E_e},
\label{eq:rap}
\end{equation}
and similarly for $D$.
The c.m.\ energy is $\sqrt S=2\sqrt{E_pE_e}$.

In a second step, the cross section of $p\overline{p}\to CD+X$ in
hadroproduction is obtained from the one of $pe^\pm\to e^\pm CD+X$ in
photoproduction by substituting $f_{\gamma/e}(x_-)=\delta(1-x_-)$ and
replacing the photon PDFs $f_{b/\gamma}(x_b,M)$ with their antiproton
counterparts $f_{b/\overline{p}}(x_b,M)$.
In hadron-collider experiments, the $p\overline{p}$ c.m.\ frame and the
laboratory one usually coincide.
The cross section of $pp\to CD+X$ in hadroproduction is accordingly evaluated
with $f_{b/p}(x_b,M)$.

We now turn to the partonic subprocesses $ab\to Q\overline{Q}[n]D$.
The differential cross section of such a process is calculated from the
pertaining transition-matrix element ${\cal T}$ as
$d\sigma/dt=\overline{|{\cal T}|^2}/(16\pi s^2)$, where the average is over
the spin and color degrees of freedom of $a$ and $b$ and the spin of $D$ is
summed over.
We apply the covariant-projector method of Ref.~\cite{pet} to implement the
$Q\overline{Q}$ Fock states $n$ according to the NRQCD factorization formalism
\cite{bbl}.

The following partonic subprocesses contribute to LO in $\alpha_s$ and $v$:
\begin{eqnarray}
\gamma\gamma&\to&Q\overline{Q}[\varsigma^{(1)}]N,
\label{eq:ppZ}\\
\gamma g&\to&Q\overline{Q}[\varsigma^{(8)}]N,
\label{eq:pgZ}\\
gg&\to&Q\overline{Q}[\varsigma^{(1)}]N,
\label{eq:ggZ}\\
gg&\to&Q\overline{Q}[\varsigma^{(8)}]N,
\label{eq:coZ}\\
q\overline{q}&\to&Q\overline{Q}[\varsigma^{(8)}]N,
\label{eq:qqZ}\\
q_u\overline{q}_d&\to&Q\overline{Q}\left[{}^3\!S_1^{(8)}\right]W^+,
\label{eq:udW}\\
q_d\overline{q}_u&\to&Q\overline{Q}\left[{}^3\!S_1^{(8)}\right]W^-,
\label{eq:duW}
\end{eqnarray}
where $q=u,d,s,c$; $q_u=u,c$; $q_d=d,s$; $N=\gamma,Z$; and
$\varsigma={}^1\!S_0,{}^3\!S_1,{}^1\!P_1,{}^3\!P_J$ with $J=0,1,2$.
For the reason explained above, $q=c$ and $q_u=c$ must not be included if
$Q=c$.
The processes $\gamma\gamma\to Q\overline{Q}[\varsigma^{(8)}]N$ and
$\gamma g\to Q\overline{Q}[\varsigma^{(1)}]N$ are forbidden by color
conservation.
Furthermore, the processes $q\overline{q}\to Q\overline{Q}[\varsigma^{(1)}]N$
are prohibited because the $Q$-quark line is connected with the $q$-quark line
by one gluon, which transmits color to the $Q\overline{Q}$ pair.
For a similar reason and due to the fact that the $W$ boson must be emitted
from the initial-state quarks if a $Q\overline{Q}$ pair is to be produced,
processes (\ref{eq:udW}) and (\ref{eq:duW}) only come with
$n={}^3\!S_1^{(8)}$.

In the Higgs-boson case $D=H$, the contributing partonic subprocesses are
analogous to Eqs.~(\ref{eq:ppZ})--(\ref{eq:qqZ}), except that, due to
charge-conjugation invariance,
$\gamma\gamma\to Q\overline{Q}[\varsigma^{(1)}]H$,
$\gamma g\to Q\overline{Q}[\varsigma^{(8)}]H$, and
$gg\to Q\overline{Q}[\varsigma^{(1)}]H$ are forbidden for
$\varsigma={}^3\!S_1,{}^1\!P_1$, and
$q\overline{q}\to Q\overline{Q}[\varsigma^{(8)}]H$ is forbidden for
$\varsigma={}^1\!S_0,{}^3\!P_J$.

The differential cross sections $d\sigma/dt$ of processes
(\ref{eq:ppZ})--(\ref{eq:duW}) are listed in the Appendix.
We combine the results proportional to the CO MEs
$\left\langle{\cal O}^{\psi(nS)}\left[{}^3\!P_J^{(8)}\right]\right\rangle$ and
$\left\langle{\cal O}^{\chi_{cJ}}\left[{}^3\!S_1^{(8)}\right]\right\rangle$
exploiting the multiplicity relations of Eq.~({\ref{eq:mul}).
We include large logarithmic corrections due to the running of the 
fine-structure constant from the Thomson limit to the electroweak scale by
expressing the gauge couplings in terms of Fermi's constant $G_F$, as
$g=2^{1/4}G_F^{1/2}m_Z$ and $g^\prime=2^{3/4}G_F^{1/2}m_W$.
Furthermore, we define the $Zq\overline{q}$ vector and axial-vector couplings
as $v_q=I_q^3-2e_q\sin^2\theta_w$ and $a_q=I_q^3$, respectively, where $e_q$
is the fractional electric charge of quark $q$, $I_q^3$ is the third component
of weak isospin of its left-handed component, and $\theta_w$ is the weak
mixing angle, which is fixed by $\cos\theta_w=m_W/m_Z$.
The results for processes (\ref{eq:pgZ}) and (\ref{eq:ggZ}) may be obtained
from those for process (\ref{eq:ppZ}) by adjusting the overall color factors
as specified in the Appendix.
The results for $N=\gamma$ emerge from the ones for $N=Z$ by adjusting the
couplings and mass of the $Z$ boson, by substituting $g=e$, $v_q=e_q$,
$a_q=0$, and $m_Z=0$, where $e=\sqrt{4\pi\alpha}$ is the electron charge
magnitude, and they are not presented separately.
They vanish for processes (\ref{eq:ppZ})--(\ref{eq:ggZ}) with
$\varsigma={}^1\!S_0,{}^3\!P_J$ and for process (\ref{eq:qqZ}) with
$\varsigma={}^1\!P_1$ because the photon has no axial-vector coupling.

In the case of $N=\gamma$, the differential cross sections $d\sigma/dt$ of
processes (\ref{eq:ppZ})--(\ref{eq:coZ}) with $\varsigma={}^1\!P_1$ and of
process (\ref{eq:qqZ}) with $\varsigma={}^3\!S_1,{}^3\!P_J$ are plagued by
infrared singularities in the limit of the final-state photon being soft,
{\it i.e.}, for $t=u=0$.
In addition, process (\ref{eq:qqZ}) with $\varsigma={}^3\!S_1$ also suffers
from a $u$- or $t$-channel singularity if the final-state photon is hard and
collinear to the initial-state $q$ or $\overline{q}$ quarks, {\it i.e.}, for
$t<u=0$ or $u<t=0$, respectively.
Owing to the identity $p_T^2=tu/s$, both the soft and collinear limits entail
$p_T\to0$.
By the same token, in the evaluations of Eq.~(\ref{eq:master}) and its
counterparts for lepton-hadron and hadron-hadron scattering, all these
singularities can be avoided by imposing a lower cut-off on $p_T$.
On the other hand, detailed inspection of these equations reveals that
$d\sigma/dp_T$ is finite in the limit $p_T\to0$ if $d\sigma/dt$ is.\footnote{%
In this sense, we disagree with the statement made in Ref.~\cite{cac} that
process (\ref{eq:coZ}) with $\varsigma={}^1\!S_0,{}^3\!P_0,{}^3\!P_2$ can
produce collinear or infrared singularities.
The triple-gluon vertex present in the contributing Feynman diagrams is
innocuous because the virtual gluon has a timelike virtuality in excess of
$m_C^2$.}
The cases $D=Z,W,H$ are devoid of such singularities because $m_D$ acts as a
regulator.

To our knowledge, the formulas presented in the Appendix cannot be found
elsewhere in the literature.
However, the literature contains analytic results for the partonic
subprocesses pertinent to the case of $C=J/\psi$ and $D=\gamma$, namely, for
processes (\ref{eq:ppZ}) \cite{ma,jap,kla}, (\ref{eq:pgZ}) \cite{meh}, and
(\ref{eq:ggZ}) \cite{meh,kim} with $\varsigma={}^3S_1$, and for processes
(\ref{eq:coZ}) and (\ref{eq:qqZ}) with $\varsigma={}^1S_0,{}^3S_1,{}^3P_J$
\cite{meh,kim}.
We agree with them, except for Eq.~(7) of Ref.~\cite{jap} (see also related
comments in Refs.~\cite{gg,kla,god}) and the term proportional to
$\left\langle O_8^{J/\psi}\left({}^3P_0\right)\right\rangle$ in Eq.~(7) of
Ref.~\cite{meh}.\footnote{%
There are two obvious typographical errors in Eq.~(8) of Ref.~\cite{meh}:
$(2m_c^2)$ and the second appearance of
$\left\langle O_8^{J/\psi}\left({}^3S_1\right)\right\rangle$ should be
replaced by $(2m_c)^2$ and
$\left\langle O_8^{J/\psi}\left({}^1S_0\right)\right\rangle$, respectively.}

\section{Numerical results}
\label{sec:three}

We are now in a position to present our numerical results.
We focus our attention on the cases $C=J/\psi,\chi_{cJ}$.
These charmonia can be efficiently identified experimentally, and their MEs
are relatively well constrained \cite{bkl}.
The predicted cross-section distributions for the $\psi^\prime$ mesons are
similar to those for the $J/\psi$ mesons, but their normalization is somewhat
suppressed due to smaller MEs \cite{bkl}.
The $\eta_c$ meson is more difficult to detect experimentally, and the $h_c$
meson is poorly known \cite{pdg}.

We first describe our theoretical input and the kinematic conditions.
We use $m_c=m_C/2=1.5$~GeV, $m_W=80.423$~GeV, $m_Z=91.1876$~GeV,
$G_F=1.16639\times10^{-5}$~GeV${}^{-2}$, $\alpha=1/137.036$, and the LO
formula for $\alpha_s^{(n_f)}(\mu)$ with $n_f=3$ active quark flavors
\cite{pdg}.
We assume $m_H=115$~GeV, a value just above the 95\%-confidence-level lower
bound on $m_H$ from direct Higgs-boson searches and in agreement with
tantalizing hints for the direct observation of the Higgs-boson signal
delivered towards the end of the LEP2 running phase.
As for the photon PDFs, we use the LO set from Gl\"uck, Reya, and Schienbein
(GRS) \cite{grs}, which is the only available one that is implemented in the
fixed-flavor-number scheme, with $n_f=3$.
As for the proton PDFs, we use the LO set from Martin, Roberts, Stirling, and
Thorne (MRST98LO) \cite{mrst}.
We adopt $\Lambda_{\rm QCD}^{(3)}=204$~MeV from Ref.~\cite{grs}, which happens
to precisely correspond to $\Lambda_{\rm QCD}^{(4)}=174$~MeV, the value
employed in Ref.~\cite{mrst}, if the matching scale is taken to be $m_c$.
We choose the renormalization and factorization scales to be
$\mu=M=m_T^C$ if $D=\gamma$ and to be $\mu=M=\sqrt{m_T^Cm_T^D}$ if $D=Z,W,H$.
As for the $J/\psi$ and $\chi_{cJ}$ MEs, we adopt the set determined in
Ref.~\cite{bkl} using the MRST98LO proton PDFs.
Specifically,
$\left\langle{\cal O}^{J/\psi}\left[{}^3\!S_1^{(1)}\right]\right\rangle$ and
$\left\langle{\cal O}^{\chi_{c0}}\left[{}^3\!P_0^{(1)}\right]\right\rangle$
were extracted from the measured partial decay widths of $J/\psi\to l^+l^-$
and $\chi_{c2}\to\gamma\gamma$ \cite{pdg}, respectively, while
$\left\langle{\cal O}^{J/\psi}\left[{}^1\!S_0^{(8)}\right]\right\rangle$,
$\left\langle{\cal O}^{J/\psi}\left[{}^3\!S_1^{(8)}\right]\right\rangle$,
$\left\langle{\cal O}^{J/\psi}\left[{}^3\!P_0^{(8)}\right]\right\rangle$, and
$\left\langle{\cal O}^{\chi_{c0}}\left[{}^3\!S_1^{(8)}\right]\right\rangle$
were fitted to the transverse-momentum distributions of $J/\psi$ and
$\chi_{cJ}$ inclusive hadroproduction \cite{abe} and the cross-section ratio
$\sigma_{\chi_{c2}}/\sigma_{\chi_{c1}}$ \cite{aff} measured at the Tevatron.
The fit results for
$\left\langle{\cal O}^{J/\psi}\left[{}^1\!S_0^{(8)}\right]\right\rangle$ and
$\left\langle{\cal O}^{J/\psi}\left[{}^3\!P_0^{(8)}\right]\right\rangle$ are
strongly correlated, so that the linear combination
\begin{equation}
M_r^{J/\psi}
=\left\langle{\cal O}^{J/\psi}\left[{}^1\!S_0^{(8)}\right]\right\rangle
+\frac{r}{m_c^2}
\left\langle{\cal O}^{J/\psi}\left[{}^3\!P_0^{(8)}\right]\right\rangle,
\label{eq:mr}
\end{equation}
with a suitable value of $r$, is quoted.
Unfortunately, Eq.~(\ref{eq:master}) is sensitive to different linear
combination of
$\left\langle{\cal O}^{J/\psi}\left[{}^1\!S_0^{(8)}\right]\right\rangle$ and
$\left\langle{\cal O}^{J/\psi}\left[{}^3\!P_0^{(8)}\right]\right\rangle$ than 
appears in Eq.~(\ref{eq:mr}).
In want of more specific information, we thus make the democratic choice
$\left\langle{\cal O}^{J/\psi}\left[{}^1\!S_0^{(8)}\right]\right\rangle
=\left(r/m_c^2\right)
\left\langle{\cal O}^{J/\psi}\left[{}^3\!P_0^{(8)}\right]\right\rangle
=M_r^{J/\psi}/2$.

We now discuss the photon flux functions that enter our predictions for
photoproduction at TESLA and THERA.
The energy spectrum of the bremsstrahlung photons is well described in the WWA
\cite{wei} by Eq.~(27) of Ref.~\cite{fri}.
We assume that the scattered electrons and positrons will be antitagged, as
was usually the case at LEP2, and take the maximum scattering angle to be
$\theta_{\rm max}=25$~mrad \cite{theta}.
The energy spectrum of the beamstrahlung photons is approximately described by
Eq.~(2.14) of Ref.~\cite{bar}.
It is controlled by the effective beamstrahlung parameter $\Upsilon$, which is
given by Eq.~(2.10) of that reference.
Inserting the relevant TESLA parameters for the $\sqrt S=500$~GeV baseline
design specified in Table~1.3.1 of Ref.~\cite{tesla} in that formula, we
obtain $\Upsilon=0.053$.
In the case of the $e^+e^-$ mode of TESLA, we coherently superimpose the WWA
and beamstrahlung spectra, while, in the case of THERA, we only use the WWA
spectrum. 
Finally, in the case of the $\gamma\gamma$ mode of TESLA, the energy spectrum
of the back-scattered laser photons is given by Eq.~(6a) of Ref.~\cite{gin}.
It depends on the parameter $\kappa=s_{e\gamma}/m_e^2-1$, where
$\sqrt{s_{e\gamma}}$ is the c.m.\ energy of the charged lepton and the laser 
photon, and it extends up to $x_{\rm max}=\kappa/(\kappa+1)$, where $x$ is the
energy of the back-scattered photons in units of $\sqrt S/2$.
The optimal value of $\kappa$ is $\kappa=2\left(1+\sqrt2\right)\approx4.83$
\cite{tel}, which we adopt; for larger values of $\kappa$, $e^+e^-$ pairs
would be created in the collisions of laser and back-scattered photons.

Since our study is at an exploratory level, we refrain from presenting a
quantitative estimate of the theoretical uncertainties in our predictions.
However, experience from previous analyses of charmonium production within the
NRQCD factorization formalism \cite{gg,ep} leads us to expect relative errors
of the order of $\pm50\%$.

\begin{table}[ht]
\begin{center}
\caption{C.m.\ energies $\protect\sqrt S$, design luminosities $L$, numbers of
dedicated experiments, and total cross sections $\sigma_1$ corresponding to a
yield of one signal event per year of operation for the considered colliders.}
\label{tab:col}
\medskip
\begin{tabular}{|lcccc|} \hline\hline
Collider & $\sqrt S$ [TeV] & $L$ [$10^{32}$~cm${}^{-2}$s${}^{-1}$] &
No.\ of experiments & $\sigma_1$ [fb] \\
\hline
TESLA $e^+e^-$ mode & 0.5 & 340 & 1 & 0.0029 \\
TESLA $\gamma\gamma$ mode & 0.5 & 60 & 1 & 0.017 \\
THERA & 1 & 0.041 & 1 & 24 \\
Tevatron Run II & 2 & 2 & 2 & 0.25 \\
LHC & 14 & 100 & 2 & 0.005 \\
\hline\hline
\end{tabular}
\end{center}
\end{table}

Table~\ref{tab:col} gives the c.m.\ energies $\sqrt S$, design luminosities
$L$, numbers of dedicated experiments, and total cross sections $\sigma_1$
corresponding to a yield of one signal event per year of operation for the
various colliders considered.
At THERA, the proton and lepton energies in the laboratory frame are planned
to be $E_p=1$~TeV and $E_e=250$~GeV \cite{thera}.
One year of operation is usually taken to be $10^7$~s, which corresponds to a
duty factor of approximately 33\% and can be achieved, {\it e.g.}, by 200 days
of running with an efficiency of 60\%.
Thus, the figures for $L$ in units of $10^{32}$~cm${}^{-2}$s${}^{-1}$
presented in Table~\ref{tab:col} equally correspond to the integrated
luminosity $\int dt\,L$ in units of fb${}^{-1}$ per year and experiment.
At the Tevatron and the LHC, the measurements can be performed simultaneously
by two dedicated experiments.
We thus obtain the values of $\sigma_1$ specified in Table~\ref{tab:col}.

We are now in a position to present our numerical results.
Figures~\ref{fig:ee}--\ref{fig:pp} are devoted to $e^+e^-\to e^+e^-CD+X$ in
the $e^+e^-$ and $\gamma\gamma$ modes of TESLA, to $pe^\pm\to e^\pm CD+X$ at
THERA, to $p\overline{p}\to CD+X$ at the Tevatron (Run~II), and to
$pp\to CD+X$ at the LHC, respectively.
In each figure, parts (a) and (b) give the $p_T$ distributions $d\sigma/dp_T$
and the $y_C$ distributions $d\sigma/dy_C$, respectively.
In each part, there are four frames, which refer to $D=\gamma,Z,W,H$,
respectively.
In each frame, we separately consider $C=J/\psi,\chi_{cJ}$, both in the CSM
and in NRQCD.
It is summed over $C=\chi_{c0},\chi_{c1},\chi_{c2}$ and $D=W^+,W^-$.
In the case of $D=\gamma$, the $y_C$ distributions are evaluated imposing the
cut $p_T>1$~GeV in order to exclude the infrared and collinear singularities
mentioned in Sec.~\ref{sec:two}.
Figure~\ref{fig:pe}(b) refers to the laboratory frame, $y_C^{\rm lab}$ being
related to its counterpart $y_C$ in the c.m.\ frame by Eq.~(\ref{eq:rap}).

We start the discussion of the figures with a few general observations.
\begin{enumerate}
\item
In all considered types of experiments, the associated production of 
$\chi_{cJ}+\gamma$, $J/\psi+W$, $\chi_{cJ}+W$, and $J/\psi+H$ is forbidden in
the CSM to the order considered.
\item
The cross section ratio of $J/\psi+W$ and $\chi_{cJ}+W$ associated production,
which proceeds through processes (\ref{eq:udW}) and (\ref{eq:duW}), is always
$\left\langle{\cal O}^{J/\psi}\left[{}^3\!S_1^{(8)}\right]\right\rangle/
\sum_{J=0}^2
\left\langle{\cal O}^{\chi_{cJ}}\left[{}^3\!S_1^{(8)}\right]\right\rangle$
${}\approx0.21$ \cite{bkl}.
\item
With increasing value of $m_{CD}$, the CO process (\ref{eq:qqZ}) with
$\varsigma={}^3\!S_1$ generally gains relative importance, since its cross
section involves a gluon propagator with small virtuality, $q^2=m_C^2$, and
is, therefore, enhanced by powers of $m_{CD}^2/m_C^2$ relative to those of the
other contributing processes.
In the fragmentation picture \cite{kra}, this cross section would be evaluated
by convoluting the cross section of $q\overline{q}\to gN$ with the
$g\to c\overline{c}\left[{}^3\!S_1^{(8)}\right]$ fragmentation function
\cite{yua}.
Processes (\ref{eq:udW}) and (\ref{eq:duW}) also benefit from this type of
enhancement.
\item
The processes with $D=\gamma$ generally have much larger cross sections than
those with $D=Z,W,H$ because the available phase space is considerably ampler
and the infrared and collinear singularities at $p_T=0$, albeit eliminated by
a minimum-$p_T$ cut, still feed into the finite parts of the cross sections.
\item
As mentioned in the Introduction, the cross sections of the processes with
$D=H$ are suppressed relative to those with $D=Z,W$ by the smallness of the
charm Yukawa coupling.
\item
As is evident from Eqs.~(\ref{eq:ppZ}) and (\ref{eq:pgZ}), the
direct-photoproduction channels in $e^+e^-$ and $pe^\pm$ collisions correspond
to pure CS and CO processes, respectively.
\item
If $m_{CD}\gg m_C$, which is the case if $p_T\gg m_C$ or $D=Z,W,H$, the
resolved channels are generally suppressed against the direct ones because,
according to Eq.~(\ref{eq:xab}), the values of $x_{a,b}$ are then close to
unity, where the photon PDFs take small values.
\item
Direct photons participate in the hard scattering with their full momenta,
while resolved ones pass on only a fraction of theirs. 
Thus, in photoproduction at THERA, the direct and resolved cross sections are
peaked in the backward and forward directions, respectively.
\end{enumerate}

We now discuss the $e^+e^-$ mode of TESLA [see Figs.~\ref{fig:ee}(a) and (b)].
Here, $J/\psi+\gamma$ associated production dominantly proceeds through the
direct CS process (\ref{eq:ppZ}) with $\varsigma={}^3\!S_1$ and $N=\gamma$.
Consequently, in the first frames of Figs.~\ref{fig:ee}(a) and (b), the dotted
and solid lines are superjacent.
$\chi_{cJ}+\gamma$ associated production only proceeds through singly-resolved
or doubly-resolved CO processes and is accordingly suppressed relative to
$J/\psi+\gamma$ associated production.
For the reasons exposed in the preceding paragraph, the reactions with $D=Z$
chiefly proceed through the direct CS processes (\ref{eq:ppZ}) with
$\varsigma={}^3\!S_1,{}^3\!P_J$ and $N=Z$.
The contributing CO processes are singly or doubly resolved and, therefore,
strongly suppressed.
The reactions with $D=W$ are mediated by the doubly-resolved CO processes
(\ref{eq:udW}) and (\ref{eq:duW}) and are accordingly suppressed.
$\chi_{cJ}+H$ associated production dominantly proceeds through direct CS 
processes, while the contributing CO processes are doubly resolved and,
therefore, heavily suppressed.
Consequently, in the fourth frames of Figs.~\ref{fig:ee}(a) and (b), the
short-dashed and medium-dashed lines are superjacent.
On the other hand, $J/\psi+H$ associated production only proceeds through
singly-resolved or doubly-resolved CO processes and is accordingly suppressed 
relative to $\chi_{cJ}+H$ associated production.

We now turn to the $\gamma\gamma$ mode of TESLA [see Figs.~\ref{fig:gg}(a) and
(b)].
In contrast to the energy spectra of bremsstrahlung and beamstrahlung, which
are strongly peaked at $x=0$, the one of the back-scattered laser photons is
evenly spread in the lower $x$ range and exhibits a maximum at
$x=x_{\rm max}$.
According to Eq.~(\ref{eq:xpm}), large values of $p_T$ or $m_D$ entail large
values of $x_\pm$.
This explains why, in the $\gamma\gamma$ mode of TESLA, the $p_T$ spectra are
less steep and the cross sections for $D=Z,W,H$ are larger than in the
$e^+e^-$ mode.
Furthermore, the influence of the singly and doubly resolved channels is
generally increased because the average photon energy is larger.
Here, the CSM prediction for $J/\psi+\gamma$ associated production is
dominated by the doubly-resolved process (\ref{eq:ggZ}) with
$\varsigma={}^3\!S_1$ and $N=\gamma$ in the lower $p_T$ range, for
$p_T\alt5$~GeV, while the direct process (\ref{eq:ppZ}) with
$\varsigma={}^3\!S_1$ and $N=\gamma$ preponderates for larger values of $p_T$.
Beholding the first frame of Fig.~\ref{fig:gg}(b), we observe that the $y_C$
distribution of $\chi_{cJ}+\gamma$ associated production is peaked in the very
forward and backward directions, close to the kinematic boundaries.
This may be traced to the single-resolved process (\ref{eq:pgZ}) with
$\varsigma={}^3\!S_1$ and $N=\gamma$.
This partonic subprocess also generates the pronounced shoulders in the $y_C$
distribution of $J/\psi+\gamma$ associated production in NRQCD, which is shown
in the same figure.
In the case of the $e^+e^-$ mode, it is suppressed by the softness of the
effective photon-energy spectrum.
The increased influence of the singly-resolved and doubly-resolved CO
processes is reflected in the first frame of Fig.~\ref{fig:gg}(a) by the
dispartment of the dotted and solid lines.
Similar separations of CSM and NRQCD results are visible in the second frames
of Figs.~\ref{fig:gg}(a) and (b) for $D=Z$.

We now move on to THERA [see Figs.~\ref{fig:pe}(a) and (b)].
In the CSM, $J/\psi+\gamma$ associated production now only happens in the
resolved channel, through process (\ref{eq:ggZ}) with $\varsigma={}^3\!S_1$
and $N=\gamma$, while, in NRQCD, it also takes place in the direct one,
through process (\ref{eq:pgZ}) with $\varsigma={}^3\!S_1$ and $N=\gamma$.
This explains why the dotted line in the first frame of Fig.~\ref{fig:pe}(b)
is peaked in the forward direction, while the solid one also exhibits a
shoulder in the backward direction.
The fact that this shoulder is not prominent may be understood by observing
that
$\left\langle{\cal O}^{J/\psi}\left[{}^3\!S_1^{(8)}\right]\right\rangle/
\left\langle{\cal O}^{J/\psi}\left[{}^3\!S_1^{(1)}\right]\right\rangle\approx
3.4\times10^{-3}$ \cite{bkl}.
Detailed inspection reveals that, contrary to na\"\i ve expectations, the
direct contribution is at least one order of magnitude suppressed against the
resolved one, even at large values of $p_T$.
From the same figure, we also glean that the CSM result almost exhausts the
resolved contribution in the lower $p_T$ range, which is decisive for the
$y_C$ distributions.
In the intermediate $p_T$ range, $6\alt p_T\alt25$~GeV, the latter is
dominated by process (\ref{eq:coZ}) with $\varsigma={}^1\!S_0,{}^3\!P_J$ and
$N=\gamma$, while, for yet larger values of $p_T$, process (\ref{eq:qqZ}) with
$\varsigma={}^3\!S_1$ and $N=\gamma$ gets in the lead.\footnote{%
This observation does not support the assertion made in Ref.~\cite{cac} that
light-quark-initiated processes are strongly suppressed and can be safely
neglected.}
In the case of $\chi_{cJ}+\gamma$ associated production, processes
(\ref{eq:pgZ}) and (\ref{eq:qqZ}) with $\varsigma={}^3\!S_1$ and $N=\gamma$
compete with each other.
The former dominates for $p_T\alt11$~GeV, while the latter prevails in the
complementary $p_T$ range.
Since the $y_C$ distributions reflect the situation in the small-$p_T$ range,
this explains why the medium-dashed line in the first frame of
Fig.~\ref{fig:pe}(b) is peaked in the backward direction.
Also in the case of $J/\psi+Z$ associated production, the direct and resolved
channels compete with each other.
The direct one is dominated by process (\ref{eq:pgZ}) with
$\varsigma={}^1\!S_0,{}^3\!P_J$ and $N=Z$, while the resolved one is dominated
by process (\ref{eq:qqZ}) with $\varsigma={}^3\!S_1$ and $N=Z$.
Here, the cross over occurs at $p_T\approx28$~GeV.
By contrast, $\chi_{cJ}+Z$ associated production is always overwhelmingly
dominated by the very same resolved process.
This explains why the solid and medium-dashed lines in the second frame of
Fig.~\ref{fig:pe}(b) are peaked in the backward and forward directions,
respectively.
In the large-$p_T$ limit, the $J/\psi+Z$ to $\chi_{cJ}+Z$ cross section ratio
approaches the same value as in the case of $D=W$ discussed under Item~2 in
the above enumeration. processes (\ref{eq:pgZ})
The CS processes (\ref{eq:ggZ}) with $\varsigma={}^3\!S_1,{}^3\!P_J$ and $N=Z$
are dramatically suppressed because they are resolved and not fragmentation
prone.
$J/\psi+H$ associated production dominantly proceeds through direct CO
processes.
On the other hand, $\chi_{cJ}+H$ associated production is only possible
through resolved processes and is, therefore, suppressed relative to
$J/\psi+H$ associated production.
There are no fragmentation-prone processes in the case $D=H$ because the Higgs
boson always couples to the charm-quark line.

We now proceed to the Tevatron [see Figs.~\ref{fig:pa}(a) and (b)]. 
For $p_T\alt5$~GeV, $J/\psi+\gamma$ associated production dominantly proceeds
through the CS process (\ref{eq:ggZ}) with $\varsigma={}^3\!S_1$ and
$N=\gamma$, due to the large value of
$\left\langle{\cal O}^{J/\psi}\left[{}^3\!S_1^{(1)}\right]\right\rangle$.
For larger values of $p_T$, the CO processes take over.
Specifically, the latter are dominated by process (\ref{eq:coZ}) with
$\varsigma={}^1\!S_0,{}^3\!P_J$ and $N=\gamma$ for $p_T\alt46$~GeV, while
process (\ref{eq:qqZ}) with $\varsigma={}^3\!S_1$ and $N=\gamma$ prevails for
larger values of $p_T$.
On the other hand, $\chi_{cJ}+\gamma$ associated production is always
dominated by the latter.
Similarly, in the case of $D=Z$, process (\ref{eq:qqZ}) with
$\varsigma={}^3\!S_1$ and $N=Z$ is always dominant.
By the same token, the $J/\psi+Z$ to $\chi_{cJ}+Z$ cross section ratio is
approximately the same as in the case of $D=W$ discussed above.
On the other hand, the CS processes (\ref{eq:ggZ}) with
$\varsigma={}^3\!S_1,{}^3\!P_J$ and $N=Z$ are not fragmentation prone and are,
therefore, vigorously suppressed.

Finally, we arrive at the LHC [see Figs.~\ref{fig:pp}(a) and (b)]. 
Here, the $p_T$ and $y_C$ distributions generally exhibit very similar shapes
as in the case of the Tevatron, while their normalizations are increased by
approximately the same amount as the value of $\sqrt S$ is.

The experimental observation of processes that are forbidden or exceedingly
suppressed in the CSM with cross sections that are compatible with the NRQCD
predictions would provide striking evidence for the NRQCD factorization
hypothesis and firmly establish the existence of CO processes in nature.
Furthermore, precise measurements of these cross sections would lead to useful
constraints on the appearing CO MEs.
On the other hand, the experimental study of processes that are dominated by
CS channels would allow for independent determinations of the CS MEs, which
can be compared with the results obtained using traditional methods, so as to
lead to valuable consistency checks.
Our analysis allows us to identify processes of all these categories.
As mentioned above, the associated production of 
$\chi_{cJ}+\gamma$, $J/\psi+W$, $\chi_{cJ}+W$, and $J/\psi+H$ exclusively
proceeds through CO processes.
At THERA, the Tevatron, and the LHC, the associated production of
$J/\psi+\gamma$ and $\chi_{cJ}+H$ with $p_T\gg m_C$ and of $J/\psi+Z$ and
$\chi_{cJ}+Z$ with arbitrary values of $p_T$ is greatly dominated by CO
processes.
On the other hand, examples where CO processes play a negligible r\^ole
include the associated production of $J/\psi+\gamma$, $J/\psi+Z$,
$\chi_{cJ}+Z$, and $\chi_{cJ}+H$ in the $e^+e^-$ mode of TESLA, of
$\chi_{cJ}+H$ in the $\gamma\gamma$ mode of TESLA, and of $J/\psi+\gamma$ and
$\chi_{cJ}+H$ with small values of $p_T$ at THERA, the Tevatron, and the LHC.

We conclude this section by assessing the observability of the various
processes in the various experiments with the aid of Table~\ref{tab:col}.
The processes with $D=\gamma$ will abundantly take place in all considered
experiments.
The processes with $D=Z,W$ will produce considerable yields at the hadron
colliders, namely, several hundred (ten thousand) events per year at the
Tevatron (LHC), while they significantly fall short of the one-event-per-year
mark at TESLA and THERA.
As expected, the processes with $D=H$ are predicted to be far too rare to be
observable in any of the considered experiments.
In turn, their observation would hint at new physics beyond the SM.

\section{Conclusions}
\label{sec:four}

We studied the associated production of heavy quarkonia $C$ with electroweak
bosons $D$ in photon-photon, photon-hadron, and hadron-hadron collisions to LO
in the NRQCD factorization formalism.
We considered all experimentally established heavy quarkonia, with
${}^{2S+1}L_J={}^1\!S_0,{}^3\!S_1,{}^1\!P_1,{}^3\!P_J$, and all electroweak
bosons of the SM, $D=\gamma,Z,W,H$.
We listed all contributing partonic cross sections, except for those with
$D=H$, which will be irrelevant for phenomenology in the foreseeable future.
We presented numerical results for any combination of $C=J/\psi,\chi_{cJ}$ and
$D=\gamma,Z,W,H$ appropriate for TESLA in the $e^+e^-$ and $\gamma\gamma$
modes, THERA, Run~II of the Tevatron, and the LHC.

At TESLA and THERA, only the processes with $D=\gamma$ will have observable
cross sections, while, at the Tevatron and the LHC, this is also the case for
$D=Z,W$.
Observation of $\chi_{cJ}+\gamma$, $J/\psi+W$, and $\chi_{cJ}+W$ associated
production, which are pure CO processes, would give strong support to the
NRQCD factorization hypothesis and, if measured with sufficient precision,
allow for independent determinations of
$\left\langle{\cal O}^{J/\psi}\left[{}^3\!S_1^{(8)}\right]\right\rangle$ and
$\left\langle{\cal O}^{\chi_{c0}}\left[{}^3\!S_1^{(8)}\right]\right\rangle$.
A similar statement applies to $J/\psi+Z$ and $\chi_{cJ}+Z$ associated
production at the Tevatron and the LHC, which are overwhelmingly dominated by
CO processes.
On the other hand, CO processes play a subordinate r\^ole for $J/\psi+\gamma$
associated production in the $e^+e^-$ mode of TESLA for arbitrary values of
$p_T$ and at THERA, the Tevatron, and the LHC for small values of $p_T$.
This offers the opportunity to extract a new value of
$\left\langle{\cal O}^{J/\psi}\left[{}^3\!S_1^{(1)}\right]\right\rangle$ and
thus to allow for a consistency check.
Alternatively, assuming
$\left\langle{\cal O}^{J/\psi}\left[{}^3\!S_1^{(1)}\right]\right\rangle$ to be
sufficiently well known, one may get a better handle on the gluon PDFs of the
photon and proton by fitting THERA, Tevatron, and LHC data of $J/\psi+\gamma$
associated production in the lower $p_T$ range.

\bigskip

\centerline{\bf Acknowledgments}

\smallskip

B.A.K. and C.P.P. are grateful to the CERN Theoretical Physics Division and
the Second Institute for Theoretical Physics of Hamburg University,
respectively, for their hospitality during visits when this paper was
prepared.
The research of B.A.K. was supported in part by the Deutsche
Forschungsgemeinschaft through Grant No.\ KN~365/1-1 and by the
Bundesministerium f\"ur Bildung und Forschung through Grant No.\ 05~HT1GUA/4.
The research of C.P.P. was supported by the Office of the Vice President for
Academic Affairs of the University of the Philippines.
The research of L.Z. was supported by the Studienstiftung des deutschen Volkes
through a PhD scholarship.

\def\theequation{\Alph{section}.\arabic{equation}}
\begin{appendix}
\setcounter{equation}{0}
\section{Partonic cross sections}

In this appendix, we list the differential cross sections $d\sigma/dt$ for
processes (\ref{eq:ppZ})--(\ref{eq:duW}) with $N=Z$ to be inserted in
Eq.~(\ref{eq:master}) and its counterparts for lepton-hadron and hadron-hadron
scattering.
The results for $N=\gamma$ are recovered as explained in Sec.~\ref{sec:two}.
Our results read

\begin{eqnarray}
\lefteqn{\frac{d\sigma}{dt}\left(\gamma\gamma\to Q\overline{Q}
\left[{}^1\!S_0^{(1)}\right]Z\right)
=\frac{1024 \pi \alpha^2 g^2 a_Q^2 (2 m_Z^2 - s - t - u)^2}{81 M m_Z^2 s^2
(m_Z^2 - s - t)^2 (m_Z^2 - s - u)^2 (2 m_Z^2 - t - u)^2}}
\nonumber\\
&&{}\times
[ m_Z^8 - 2 m_Z^6 (t + u)
- m_Z^4 (s^2 - t^2 - 4 t u - u^2) - 2 m_Z^2 t u (t + u) + t^2 u^2] ,
\end{eqnarray}

\begin{eqnarray}
\lefteqn{\frac{d\sigma}{dt}\left(\gamma\gamma\to Q\overline{Q}
\left[{}^3\!S_1^{(1)}\right]Z\right)
=\frac{-1024 \pi \alpha^2 g^2 v_Q^2}{243 M s^2 (m_Z^2 - s - t)^2
(m_Z^2 - s - u)^2 (2 m_Z^2 - t - u)^2}}
\nonumber\\
&&{} \times \{ m_Z^{10} - 4 m_Z^8 (3 s + t + u) + 
  m_Z^6 [22 s^2 + 26 s (t + u)+ 5 t^2 + 12 t u + 5 u^2]
\nonumber\\
&&{} - 2 m_Z^4 [5 s^3
 + 14 s^2 (t + u) + s (8 t^2 + 23 t
  u + 8 u^2) + (t + u) (t^2 + 5 t u + u^2)]
\nonumber\\
&&{}  -  m_Z^2 [s^4 - 4 s^3 (t + u) -  s^2 (9 t^2 + 26 t u + 9 u^2) - 2 s (t +
 u) (t^2 + 10 t u + u^2)
\nonumber\\
&&{}  - t u (4 t^2 + 9
  t u + 4 u^2)]
\nonumber\\
&&{}  - 
  2 [s^3 (t^2 + t u + u^2) + s^2 (t + u)^3  
  +  s t u (t^2 + 3 t u + u^2) + t^2 u^2 (t + u)]\} ,
\end{eqnarray}

\begin{eqnarray}
\lefteqn{\frac{d\sigma}{dt}\left(\gamma\gamma\to Q\overline{Q}
\left[{}^1\!P_1^{(1)}\right]Z\right)
=\frac{-8192 \pi \alpha^2 g^2 v_Q^2}{243 M^3 s^2 (m_Z^2 - s - t)^3
(m_Z^2 - s - u)^3 (2 m_Z^2 - t - u)^4}}
\nonumber\\
&&{} \times \{ 48 m_Z^{18} - 16 m_Z^{16} [11 s + 15 (t + u)]
\nonumber\\
&&{}  + 
  2 m_Z^{14} [141 s^2 + 374 s (t + u) + 4 (65 t^2 + 134 t u +
  65 u^2)]
\nonumber\\
&&{}  -  4 m_Z^{12} [65 s^3 + 255 s^2 (t + u) + s (341 t^2
  + 696 t u + 341 u^2)
\nonumber\\
&&{}  +  32 (t + u) (5 t^2 + 11 t u + 5 u^2)]
\nonumber\\
&&{}  + m_Z^{10} [138 s^4  + 798 s^3
  (t + u) +  10 s^2 (158 t^2 + 315 t u + 158 u^2)
\nonumber\\
&&{}  +  s (t + u) (1399 t^2 + 2958 t u + 1399 u^2)
\nonumber\\
&&{}   + 491 t^4 +
  2188 t^3 u +  3378 t^2 u^2 + 2188 t u^3 + 491 u^4]
\nonumber\\
&&{}  -  m_Z^8 [24 s^5 + 356 s^4 (t +
  u) +  20 s^3 (53 t^2 + 101 t u + 53 u^2)
\nonumber\\
&&{}  + 10 s^2 (t + u) (138 t^2 + 265 t u + 138 u^2) + s (t + u)^2 
  (887 t^2 + 1956 t u + 887 u^2)
\nonumber\\
&&{}  + (t + u) (241 t^4 + 1188 t^3 u + 1846 t^2 u^2 + 1188 t u^3 + 241 u^4)]
\nonumber\\
&&{}  -  m_Z^6 [18 s^6
  - 60 s^5 (t + u) - s^4 (417 t^2 + 734 t u + 417 u^2)
\nonumber\\
&&{}   -  s^3
  (t + u) (801 t^2 + 1298 t u + 801 u^2)
\nonumber\\
&&{}  -  s^2 (749 t^4 + 2772 t^3 u + 4068 t^2 u^2 + 2772 t
  u^3 + 749 u^4)
\nonumber\\
&&{}  -  s (t + u) (359 t^4 + 1534 t^3 u + 2362 t^2 u^2 + 1534 t u^3 + 359 u^4)
\nonumber\\
&&{}  -  2 (t +
  u)^2 (37 t^4+ 217 t^3 u  + 332 t^2 u^2 + 217 t u^3 + 37
  u^4)]
\nonumber\\
&&{}   + m_Z^4 [12 s^7 +
  16 s^6 (t + u) -  s^5 (83 t^2 + 122 t u + 83 u^2)
\nonumber\\
&&{}  - s^4
  (t + u) (275 t^2 + 356 t u + 275 u^2)
\nonumber\\
&&{} -
  s^3 (367 t^4 + 1153 t^3 u + 1580 t^2 u^2  + 1153 t u^3 + 
  367 u^4)
\nonumber\\
&&{} - s^2 (t + u) (258 t^4 + 843 t^3 u + 1214 t^2 u^2 + 843 t u^3 + 258 u^4)
\nonumber\\
&&{}  - 4 s (23 t^6 + 143 t^5 u
  + 371 t^4 u^2 +  500 t^3 u^3 + 371 t^2 u^4 + 143 t u^5 + 23
  u^6)
\nonumber\\
&&{}  - (t + u)^3
  (13 t^4 + 104 t^3 u + 150 t^2 u^2 + 104 t u^3 +  13 u^4)]
\nonumber\\
&&{}  -  m_Z^2 [2 s^8 + 6
  s^7 (t + u) - s^6 (7 t^2 + 4 t u + 7 u^2) -  52 s^5 (t + u) (t^2 + t u + u^2)
\nonumber\\
&&{}  -  s^4 (97 t^4 + 269 t^3 u + 346
  t^2 u^2 + 269 t u^3 + 97 u^4)
\nonumber\\
&&{}  -  2 s^3 (t + u) (48 t^4 + 122 t^3 u + 155 t^2 u^2 + 122 t u^3 + 48 u^4)
\nonumber\\
&&{}  - s^2
  (53 t^6 + 244 t^5 u + 531 t^4 u^2 + 674 t^3 u^3 +  531 t^2
  u^4 + 244 t u^5 + 53 u^6)
\nonumber\\
&&{}  -  s (t + u) (14 t^6 + 82 t^5 u + 217 t^4 u^2 + 290 t^3 u^3 + 217 t^2
 u^4 + 
  82 t u^5 + 14 u^6)
\nonumber\\
&&{}  -  (t + u)^4 (t^4 + 15 t^3 u + 19
  t^2 u^2 + 15 t u^3 + u^4)]
\nonumber\\
&&{}  -  (t + u) [s^7 (t + u) +  s^6 (5 t^2 + 6 t u + 5 u^2) + s^5 (t + u) (11
 t^2 + 8 t u + 11 u^2)
\nonumber\\
&&{}  + 2 s^4 (7 t^4 + 16 t^3 u + 19 t^2
  u^2 + 16 t u^3 + 7 u^4)
\nonumber\\
&&{}  +  s^3 (t + u) (11 t^4 + 21 t^3 u + 25 t^2 u^2 + 21 t u^3 + 11 u^4)
\nonumber\\
&&{}  +  s^2 (5 t^6 + 19
  t^5 u + 38 t^4 u^2 + 46 t^3 u^3 + 38 t^2 u^4 +  19 t u^5 +
  5 u^6)
\nonumber\\
&&{}  + s (t + u) (t^6 + 5 t^5 u + 14 t^4 u^2 + 18 t^3 u^3 + 14 t^2 u^4 + 5 t
 u^5 + u^6)
\nonumber\\
&&{}  + t u (t + u)^4 (t^2 + t u +
  u^2)]\} ,
\end{eqnarray}

\begin{eqnarray}
\lefteqn{\frac{d\sigma}{dt}\left(\gamma\gamma\to Q\overline{Q}
\left[{}^3\!P_J^{(1)}\right]Z\right)
=\frac{2048 \pi \alpha^2 g^2 a_Q^2}{1215 M^3 m_Z^2 s^2 (m_Z^2 - s - t)^4
(m_Z^2 - s - u)^4 (2 m_Z^2 - t - u)^4}}
\nonumber\\
&&{}\times F_J, \qquad J=0,1,2,\hspace{10cm}
\end{eqnarray}

\begin{eqnarray}
F_0
&=& 10 (2 m_Z^2 - s - t - u)^2 \{ 4 m_Z^{18} s - 4 m_Z^{16} [2 s^2 + 4 s
(t + u) - (t - u)^2]
\nonumber\\
&&{}  - 
   2 m_Z^{14} [10 s^3 - s (9 t^2 + 38 t u + 9
   u^2) + 10 (t + u) (t - u)^2]
\nonumber\\
&&{}  + 
   m_Z^{12} [80 s^4 + 176 s^3 (t + u) + 
     2 s^2 (35 t^2 + 46 t u + 35 u^2)
\nonumber\\
&&{} + 2 s (t + u) (3 t^2 - 62 t u + 3 u^2) + (t - u)^2 (41 t^2 + 90 t u + 41
 u^2)]
\nonumber\\
&&{}  - 2 m_Z^{10} [50 s^5 + 192 s^4 (t + u) + 
     8 s^3 (27 t^2 + 50 t u + 27 u^2)
\nonumber\\
&&{} + 2 s^2 (t + u) (35 t^2 + 48 t u + 35 u^2) + 
     s (13 t^4 - 23 t^3 u - 120 t^2 u^2 - 23 t u^3 + 13 u^4)
\nonumber\\
&&{} + 2 (t - u)^2 (t + u) (11 t^2 + 30 t u + 11 u^2)]
\nonumber\\
&&{}  + 
   m_Z^8 [56 s^6 + 336 s^5 (t + u) + 64 s^4 (10 t^2 + 19 t u + 10 u^2)
\nonumber\\
&&{} + 
     4 s^3 (t + u) (123 t^2 + 220 t u + 123 u^2)
\nonumber\\
&&{} + 
     s^2 (125 t^4 + 476 t^3 u + 598 t^2 u^2 + 476 t u^3 + 125 u^4)
\nonumber\\
&&{} + 
     2 s (t + u) (9 t^4 + 18 t^3 u - 110 t^2 u^2 + 18 t u^3 + 9 u^4)
\nonumber\\
&&{} + 
     2 (t - u)^2 (13 t^4 + 77 t^3 u + 130 t^2 u^2 + 77 t u^3 + 13 u^4)]
\nonumber\\
&&{}  - 
   2 m_Z^6 [6 s^7 + 64 s^6 (t + u) + s^5 (191 t^2 + 370 t u + 191 u^2)
\nonumber\\
&&{} + 
     2 s^4 (t + u) (127 t^2 + 232 t u + 127 u^2)
\nonumber\\
&&{} + 
     s^3 (157 t^4 + 582 t^3 u + 828 t^2 u^2 + 582 t u^3 + 157 u^4)
\nonumber\\
&&{} + 2 s^2 (t + u) (17 t^4 + 70 t^3 u + 70 t^2 u^2 + 70 t u^3 + 17 u^4)
\nonumber\\
&&{} + s (2 t^6 + 30 t^5 u - 9 t^4 u^2 - 102 t^3 u^3 - 
       9 t^2 u^4 + 30 t u^5 + 2 u^6)
\nonumber\\
&&{} + 
     (t - u)^2 (t + u) (4 t^4 + 35 t^3 u + 68 t^2 u^2 + 35 t u^3 + 4 u^4)]
\nonumber\\
&&{}  + 
   m_Z^4 [16 s^7 (t + u) + 86 s^6 (t + u)^2 + 2 s^5 (t + u) (95 t^2 + 182 t u
 + 95 u^2)
\nonumber\\
&&{} + s^4 (t + u)^2 
      (217 t^2 + 358 t u + 217 u^2)
\nonumber\\
&&{} + 2 s^3 (t + u) (63 t^4 + 204 t^3 u + 278 t^2 u^2 + 204 t u^3 + 63 u^4)
\nonumber\\
&&{} + 
     2 s^2 (14 t^6 + 71 t^5 u + 145 t^4 u^2 + 160 t^3 u^3 + 145 t^2 u^4 + 
       71 t u^5 + 14 u^6)
\nonumber\\
&&{} + 2 s t u (t + u) (10 t^4 + 11 t^3 u - 50 t^2 u^2 + 11 t u^3 + 10 u^4)
\nonumber\\
&&{} + (t - u)^2 (t + u)^2 (t^4 + 18 t^3 u + 47 t^2 u^2 + 18 t u^3 + u^4)]
\nonumber\\
&&{}  - 2 m_Z^2 [2 s^7 (t + u)^2 + 10 s^6 (t + u)^3 + s^5 (22 t^4 + 81 t^3 u +
 120 t^2 u^2 + 
       81 t u^3 + 22 u^4)
\nonumber\\
&&{} + 2 s^4 (t + u) (13 t^4 + 40 t^3 u + 56 t^2 u^2 + 40 t u^3 + 13 u^4)
\nonumber\\
&&{} + 4 s^3 (t^2 + t u + u^2)^2 
      (4 t^2 + 9 t u + 4 u^2)
\nonumber\\
&&{} + 
     s^2 (t + u) (4 t^6 + 13 t^5 u + 26 t^4 u^2 + 22 t^3 u^3 + 26 t^2 u^4 + 
  13 t u^5 + 4 u^6)
\nonumber\\
&&{} + s t u (t^6 + 9 t^5 u - 2 t^4 u^2 - 
       18 t^3 u^3 - 2 t^2 u^4 + 9 t u^5 + u^6)
\nonumber\\
&&{} + 
     t u (t - u)^2 (t + u)^3 (t^2 + 5 t u + u^2)]
\nonumber\\
&&{} + 
  (s + t + u)^2 (t + u)^2 [s^4 (t + u)^2 + 2 s^3 (t + u) (t^2 + u^2)
\nonumber\\
&&{} + 
     s^2 (t^2 + u^2)^2 + t^2 u^2 (t - u)^2] \} ,
\end{eqnarray}

\begin{eqnarray}
F_1
&=& -5 \{ 176 m_Z^{24} - 16 m_Z^{22} [64 s + 63 (t + u)]
\nonumber\\
&&{} + 
   12 m_Z^{20} [208 s^2 + 406 s (t + u) + 203 t^2 + 446 t u + 203 u^2]
\nonumber\\
&&{} - 
   4 m_Z^{18} [864 s^3 + 
     2341 s^2 (t + u) + 4 s (573 t^2 + 1315 t u + 573 u^2)
\nonumber\\
&&{} + 2 (t + u) (397 t^2 + 1086 t u + 397 u^2)]
\nonumber\\
&&{} + 
   2 m_Z^{16} [1700 s^4 + 4810 s^3 (t + u) + 2 s^2 (2969 t^2 + 7482 t u + 2969
 u^2)
\nonumber\\
&&{} + 
     s (t + u) (3947 t^2 + 13458 t u + 3947 u^2)
\nonumber\\
&&{} + 1157 t^4 + 6964 t^3 u + 11838 t^2 u^2 + 6964 t u^3 + 
     1157 u^4]
\nonumber\\
&&{} - 
   4 m_Z^{14} [764 s^5 + 1843 s^4 (t + u) + 
     s^3 (1409 t^2 + 4728 t u + 1409 u^2)
\nonumber\\
&&{} + 
     19 s^2 (t + u) (25 t^2 + 324 t u + 25 u^2)
\nonumber\\
&&{} + 
     s (341 t^4 + 5925 t^3 u + 11740 t^2 u^2 + 5925 t u^3 + 341 u^4)
\nonumber\\
&&{} + 
     3 (t + u) (65 t^4 + 664 t^3 u + 1350 t^2 u^2 + 664 t u^3 + 65 u^4)]
\nonumber\\
&&{} + m_Z^{12} [2552 s^6 + 6668 s^5 (t + u) + 
     4 s^4 (467 t^2 + 2408 t u + 467 u^2)
\nonumber\\
&&{} - 
     2 s^3 (t + u) (4693 t^2 - 714 t u + 4693 u^2)
\nonumber\\
&&{} - 
     s^2 (10469 t^4 + 15252 t^3 u + 5214 t^2 u^2 + 15252 t u^3 + 10469 u^4)
\nonumber\\
&&{} - 
     2 s (t + u) (1719 t^4 - 461 t^3 u - 8172 t^2 u^2 - 461 t u^3 + 1719 u^4)
\nonumber\\
&&{} - 95 t^6 + 2048 t^5 u + 11983 t^4 u^2 + 
     19840 t^3 u^3 + 11983 t^2 u^4 + 2048 t u^5 - 95 u^6]
\nonumber\\
&&{} - 2 m_Z^{10} [784 s^7 + 
     2838 s^6 (t + u) + 2 s^5 (1149 t^2 + 3098 t u + 1149 u^2)
\nonumber\\
&&{} - 
     2 s^4 (t + u) (1943 t^2 + 1008 t u + 1943 u^2)
\nonumber\\
&&{} - 
     s^3 (8747 t^4 + 25789 t^3 u + 31940 t^2 u^2 + 25789 t u^3 + 8747 u^4)
\nonumber\\
&&{} - 
     s^2 (t + u) (6261 t^4 + 17595 t^3 u + 16888 t^2 u^2 + 17595 t u^3 + 6261
 u^4)
\nonumber\\
&&{} - s (1741 t^6 + 7835 t^5 u + 10749 t^4 u^2 + 
       8966 t^3 u^3 + 10749 t^2 u^4 + 7835 t u^5 + 1741 u^6)
\nonumber\\
&&{} - (t + u) (99 t^6 + 343 t^5 u - 679 t^4 u^2 - 2118 t^3 u^3 - 679 t^2 u^4 + 
   343 t u^5 + 99 u^6)]
\nonumber\\
&&{} + 
   m_Z^8 [584 s^8 + 2956 s^7 (t + u) + 4 s^6 (1259 t^2 + 2842 t u + 
       1259 u^2)
\nonumber\\
&&{} + 2 s^5 (t + u) (237 t^2 + 2626 t u + 237 u^2)
\nonumber\\
&&{} - 
     s^4 (9273 t^4 + 30930 t^3 u + 40826 t^2 u^2 + 30930 t u^3 + 9273 u^4)
\nonumber\\
&&{} - 
     2 s^3 (t + u) (6289 t^4 + 22639 t^3 u + 28320 t^2 u^2 + 22639 t u^3 + 
  6289 u^4)
\nonumber\\
&&{} - s^2 (7039 t^6 + 40430 t^5 u + 
       86805 t^4 u^2 + 105964 t^3 u^3
\nonumber\\
&&{} + 86805 t^2 u^4 + 40430 t u^5 + 
       7039 u^6)
\nonumber\\
&&{} - 2 s (t + u) (811 t^6 + 4992 t^5 u + 9574 t^4 u^2 + 9982 t^3 u^3 + 
  9574 t^2 u^4 + 4992 t u^5 + 811 u^6)
\nonumber\\
&&{} - 2 (39 t^8 + 404 t^7 u + 1209 t^6 u^2 + 1488 t^5 u^3 + 
       1280 t^4 u^4
\nonumber\\
&&{} + 1488 t^3 u^5 + 1209 t^2 u^6 + 404 t u^7 + 39 u^8)]
\nonumber\\
&&{} - 
   2 m_Z^6 [56 s^9 + 386 s^8 (t + u) + 2 s^7 (539 t^2 + 1164 t u + 539 u^2)
\nonumber\\
&&{} + 
     2 s^6 (t + u) (595 t^2 + 1466 t u + 595 u^2)
\nonumber\\
&&{} - 
     2 s^5 (201 t^4 + 599 t^3 u + 588 t^2 u^2 + 599 t u^3 + 201 u^4)
\nonumber\\
&&{} - 
     2 s^4 (t + u) (1189 t^4 + 4965 t^3 u + 6653 t^2 u^2 + 4965 t u^3 + 1189
 u^4)
\nonumber\\
&&{} - s^3 (2455 t^6 + 15774 t^5 u + 37813 t^4 u^2 + 
       48736 t^3 u^3 + 37813 t^2 u^4 + 15774 t u^5 + 2455 u^6)
\nonumber\\
&&{} - 
     s^2 (t + u) (1127 t^6 + 7744 t^5 u + 18069 t^4 u^2 + 22220 t^3 u^3
\nonumber\\
&&{}  + 
  18069 t^2 u^4 + 7744 t u^5 + 1127 u^6)
\nonumber\\
&&{} - 
     s (213 t^8 + 2150 t^7 u + 7654 t^6 u^2 + 14044 t^5 u^3 + 16614 t^4 u^4
\nonumber\\
&&{} + 
       14044 t^3 u^5 + 7654 t^2 u^6 + 2150 t u^7 + 213 u^8)
\nonumber\\
&&{} - (t + u) (7 t^8 + 118 t^7 u + 528 t^6 u^2 + 978 t^5 u^3 + 1098 t^4 u^4
\nonumber\\
&&{}  + 
   978 t^3 u^5 + 528 t^2 u^6 + 118 t u^7 + 7 u^8)]
\nonumber\\
&&{} + 
   m_Z^4 [8 s^{10} + 76 s^9 (t + u) + 8 s^8 (44 t^2 + 93 t u + 44 u^2)
\nonumber\\
&&{} + 
     6 s^7 (t + u) (127 t^2 + 278 t u + 127 u^2)
\nonumber\\
&&{} + 
     s^6 (597 t^4 + 2324 t^3 u + 3574 t^2 u^2 + 2324 t u^3 + 597 u^4)
\nonumber\\
&&{} - 
     2 s^5 (t + u) (233 t^4 + 1377 t^3 u + 1950 t^2 u^2 + 1377 t u^3 + 233 u^4)
\nonumber\\
&&{} - s^4 (1327 t^6 + 9644 t^5 u + 25033 t^4 u^2 + 
       33280 t^3 u^3 + 25033 t^2 u^4 + 9644 t u^5 + 1327 u^6)
\nonumber\\
&&{} - 
     2 s^3 (t + u) (547 t^6 + 4136 t^5 u + 10580 t^4 u^2 + 13738 t^3 u^3
\nonumber\\
&&{}  + 
  10580 t^2 u^4 + 4136 t u^5 + 547 u^6)
\nonumber\\
&&{} - 
     s^2 (413 t^8 + 4348 t^7 u + 16425 t^6 u^2 + 32576 t^5 u^3 + 
       40108 t^4 u^4
\nonumber\\
&&{} + 32576 t^3 u^5 + 16425 t^2 u^6 + 4348 t u^7 + 
       413 u^8)
\nonumber\\
&&{} - 2 s (t + u) (31 t^8 + 409 t^7 u + 1674 t^6 u^2 + 3332 t^5 u^3 + 
  4016 t^4 u^4
\nonumber\\
&&{}  + 3332 t^3 u^5 + 1674 t^2 u^6 + 409 t u^7 + 31 u^8)
\nonumber\\
&&{} - 
     (t + u)^2 (t^8 + 34 t^7 u + 225 t^6 u^2 + 524 t^5 u^3 + 628 t^4 u^4
\nonumber\\
&&{} + 
       524 t^3 u^5 + 225 t^2 u^6 + 34 t u^7 + u^8)]
\nonumber\\
&&{} - 
   2 m_Z^2 [6 s^9 (t + u)^2 + 30 s^8 (t + u)^3 + 
     s^7 (t + u)^2 (51 t^2 + 83 t u + 51 u^2)
\nonumber\\
&&{} + 
     s^6 (t + u) (13 t^4 - 39 t^3 u - 98 t^2 u^2 - 39 t u^3 + 13 u^4)
\nonumber\\
&&{} - s^5 (70 t^6 + 613 t^5 u + 
       1782 t^4 u^2 + 2470 t^3 u^3 + 1782 t^2 u^4 + 613 t u^5 + 70 u^6)
\nonumber\\
&&{} - 
     s^4 (t + u) (104 t^6 + 873 t^5 u + 2434 t^4 u^2 + 3298 t^3 u^3 + 
  2434 t^2 u^4 + 873 t u^5 + 104 u^6)
\nonumber\\
&&{} - 
     s^3 (65 t^8 + 733 t^7 u + 2918 t^6 u^2 + 6080 t^5 u^3 + 7652 t^4 u^4
\nonumber\\
&&{} + 
       6080 t^3 u^5 + 2918 t^2 u^6 + 733 t u^7 + 65 u^8)
\nonumber\\
&&{} - 
     s^2 (t + u) (19 t^8 + 263 t^7 u + 1082 t^6 u^2 + 2252 t^5 u^3 + 
  2802 t^4 u^4
\nonumber\\
&&{}  + 2252 t^3 u^5 + 1082 t^2 u^6 + 263 t u^7 + 19 u^8)
\nonumber\\
&&{} - 
     s (t + u)^2 (2 t^8 + 43 t^7 u + 200 t^6 u^2 + 440 t^5 u^3 + 
       540 t^4 u^4
\nonumber\\
&&{} + 440 t^3 u^5 + 200 t^2 u^6 + 43 t u^7 + 2 u^8)
\nonumber\\
&&{} - t u (t + u)^3 (t^2 + t u + u^2) (t^4 + 11 t^3 u + 24 t^2 u^2 + 11 t u^3
 + 
   u^4)]
\nonumber\\
&&{} - 
   (t + u) (s + t + u)^2 [
     s^6 (t + u) (t^2 + 4 t u + u^2)
\nonumber\\
&&{} + 
     2 s^5 (2 t^4 + 11 t^3 u + 20 t^2 u^2 + 11 t u^3 + 2 u^4)
\nonumber\\
&&{} + 
     2 s^4 (t + u) (3 t^4 + 17 t^3 u + 32 t^2 u^2 + 17 t u^3 + 3 u^4)
\nonumber\\
&&{} + 2 s^3 (t + u)^2 
      (2 t^4 + 15 t^3 u + 24 t^2 u^2 + 15 t u^3 + 2 u^4)
\nonumber\\
&&{} + s^2 (t + u) (t^6 + 18 t^5 u + 48 t^4 u^2 + 68 t^3 u^3 + 48 t^2 u^4 + 
  18 t u^5 + u^6)
\nonumber\\
&&{} + 
     2 s t u (t^2 + t u + u^2) (2 t^4 + 7 t^3 u + 12 t^2 u^2 + 7 t u^3 + 2 u^4)
 + t^2 u^2 (t + u)^5]\} ,  
\end{eqnarray}

\begin{eqnarray}
F_2
&=& 96 m_Z^{24} - 64 m_Z^{22} [11 s + 15 (t + u)]
\nonumber\\
&&{} + 
   8 m_Z^{20} [192 s^2 + 713 s (t + u) + 28 (17 t^2 + 35 t u + 17 u^2)]
\nonumber\\
&&{} + 
   4 m_Z^{18} [40 s^3 - 2689 s^2 (t + u) - 2 s (2359 t^2 + 4960 t u + 
       2359 u^2)
\nonumber\\
&&{} - 12 (t + u) (173 t^2 + 384 t u + 173 u^2)]
\nonumber\\
&&{} - 
   2 m_Z^{16} [3020 s^4 - 610 s^3 (t + u) - 
     2 s^2 (7615 t^2 + 16398 t u + 7615 u^2)
\nonumber\\
&&{} - 
     3 s (t + u) (5775 t^2 + 13402 t u + 5775 u^2)
\nonumber\\
&&{} - 5647 t^4 - 25540 t^3 u - 39866 t^2 u^2 - 
     25540 t u^3 - 5647 u^4]
\nonumber\\
&&{} + 
   4 m_Z^{14} [2836 s^5 + 5845 s^4 (t + u) - 
     2 s^3 (730 t^2 + 1927 t u + 730 u^2)
\nonumber\\
&&{} - 
     2 s^2 (t + u) (5965 t^2 + 14477 t u + 5965 u^2)
\nonumber\\
&&{} - 
     s (9956 t^4 + 46445 t^3 u + 73390 t^2 u^2 + 46445 t u^3 + 9956 u^4)
\nonumber\\
&&{} - 
     2 (t + u) (1269 t^4 + 6380 t^3 u + 10334 t^2 u^2 + 6380 t u^3 + 1269 u^4)]
\nonumber\\
&&{} - m_Z^{12} [10600 s^6 + 36980 s^5 (t + u) + 4 s^4 (9559 t^2 + 19444 t u +
 9559 u^2)
\nonumber\\
&&{} - 
     2 s^3 (t + u) (5559 t^2 + 16706 t u + 5559 u^2)
\nonumber\\
&&{} - 
     s^2 (47103 t^4 + 224004 t^3 u + 357674 t^2 u^2 + 224004 t u^3 + 
       47103 u^4)
\nonumber\\
&&{} - 2 s (t + u) (15175 t^4 + 78107 t^3 u + 129108 t^2 u^2 + 78107 t u^3 + 
  15175 u^4)
\nonumber\\
&&{} - 6165 t^6 - 48348 t^5 u - 
     139987 t^4 u^2 - 195640 t^3 u^3
\nonumber\\
&&{} - 139987 t^2 u^4 - 48348 t u^5 - 
     6165 u^6]
\nonumber\\
&&{} + 
   2 m_Z^{10} [2880 s^7 + 
     13554 s^6 (t + u) + 4 s^5 (6177 t^2 + 12874 t u + 6177 u^2)
\nonumber\\
&&{} + 
     2 s^4 (t + u) (8249 t^2 + 18092 t u + 8249 u^2)
\nonumber\\
&&{} - 
     s^3 (6683 t^4 + 31527 t^3 u + 51696 t^2 u^2 + 31527 t u^3 + 6683 u^4)
\nonumber\\
&&{} - 
     s^2 (t + u) (15723 t^4 + 78965 t^3 u + 132576 t^2 u^2 + 78965 t u^3 + 
  15723 u^4)
\nonumber\\
&&{} - 3 s (2621 t^6 + 20411 t^5 u + 
       59773 t^4 u^2 + 84038 t^3 u^3
\nonumber\\
&&{} + 59773 t^2 u^4 + 20411 t u^5 + 
       2621 u^6)
\nonumber\\
&&{} - (t + u) (1257 t^6 + 11553 t^5 u + 36251 t^4 u^2 + 51926 t^3 u^3
\nonumber\\
&&{}  + 
   36251 t^2 u^4 + 11553 t u^5 + 1257 u^6)]
\nonumber\\
&&{} - 
   m_Z^8 [1864 s^8 + 10820 s^7 (t + u) + 4 s^6 (6847 t^2 + 14418 t u + 
       6847 u^2)
\nonumber\\
&&{} + 2 s^5 (t + u) (17033 t^2 + 39274 t u + 17033 u^2)
\nonumber\\
&&{} + s^4 (14141 t^4 + 71138 t^3 u + 111826 t^2 u^2 + 
       71138 t u^3 + 14141 u^4)
\nonumber\\
&&{} - 2 s^3 (t + u) (5695 t^4 + 21723 t^3 u + 37444 t^2 u^2 + 21723 t u^3 + 
  5695 u^4)
\nonumber\\
&&{} - 
     s^2 (14753 t^6 + 104550 t^5 u + 300227 t^4 u^2 + 421692 t^3 u^3
\nonumber\\
&&{} + 
       300227 t^2 u^4 + 104550 t u^5 + 14753 u^6)
\nonumber\\
&&{} - 
     2 s (t + u) (2771 t^6 + 23944 t^5 u + 75106 t^4 u^2 + 108454 t^3 u^3
\nonumber\\
&&{}  + 
  75106 t^2 u^4 + 23944 t u^5 + 2771 u^6)
\nonumber\\
&&{} - 
     2 (t + u)^2 (331 t^6 + 3842 t^5 u + 13494 t^4 u^2 + 19926 t^3 u^3
\nonumber\\
&&{} + 
       13494 t^2 u^4 + 3842 t u^5 + 331 u^6)]
\nonumber\\
&&{} + 
   2 m_Z^6 [168 s^9 + 1158 s^8 (t + u) + 12 s^7 (316 t^2 + 671 t u + 
       316 u^2)
\nonumber\\
&&{} + 4 s^6 (t + u) (1693 t^2 + 3990 t u + 1693 u^2)
\nonumber\\
&&{} + 
     s^5 (5796 t^4 + 29542 t^3 u + 47188 t^2 u^2 + 29542 t u^3 + 5796 u^4)
\nonumber\\
&&{} + 
     2 s^4 (t + u) (162 t^4 + 4996 t^3 u + 8301 t^2 u^2 + 4996 t u^3 + 162 u^4)
\nonumber\\
&&{} - 
     s^3 (3347 t^6 + 16010 t^5 u + 39319 t^4 u^2 + 53548 t^3 u^3
\nonumber\\
&&{} + 
       39319 t^2 u^4 + 16010 t u^5 + 3347 u^6)
\nonumber\\
&&{} - 
     s^2 (t + u) (2439 t^6 + 16826 t^5 u + 49619 t^4 u^2 + 71420 t^3 u^3
\nonumber\\
&&{} + 
  49619 t^2 u^4 + 16826 t u^5 + 2439 u^6)
\nonumber\\
&&{} - 
     s (645 t^8 + 7666 t^7 u + 35040 t^6 u^2 + 82140 t^5 u^3 + 
       108250 t^4 u^4
\nonumber\\
&&{} + 82140 t^3 u^5 + 35040 t^2 u^6 + 7666 t u^7 + 
       645 u^8)
\nonumber\\
&&{} - (t + u)^3 (51 t^6 + 852 t^5 u + 3539 t^4 u^2 + 
       5396 t^3 u^3 + 3539 t^2 u^4 + 852 t u^5 + 51 u^6)]
\nonumber\\
&&{} - 
   m_Z^4 [24 s^{10} + 212 s^9 (t + u) + 4 s^8 (241 t^2 + 512 t u + 241 u^2)
\nonumber\\
&&{} + 
     2 s^7 (t + u) (1179 t^2 + 2822 t u + 1179 u^2)
\nonumber\\
&&{} + 
     s^6 (2855 t^4 + 15364 t^3 u + 24930 t^2 u^2 + 15364 t u^3 + 2855 u^4)
\nonumber\\
&&{} + 
     2 s^5 (t + u) (367 t^4 + 6085 t^3 u + 10618 t^2 u^2 + 6085 t u^3 + 367 u^4)
\nonumber\\
&&{} - 
     s^4 (2117 t^6 + 1876 t^5 u - 6829 t^4 u^2 - 13152 t^3 u^3 - 
       6829 t^2 u^4 + 1876 t u^5 + 2117 u^6)
\nonumber\\
&&{} - 
     2 s^3 (t + u) (1241 t^6 + 4958 t^5 u + 11056 t^4 u^2 + 15114 t^3 u^3
\nonumber\\
&&{} + 
  11056 t^2 u^4 + 4958 t u^5 + 1241 u^6)
\nonumber\\
&&{} - 
     s^2 (1079 t^8 + 9328 t^7 u + 36475 t^6 u^2 + 80988 t^5 u^3 + 
       105492 t^4 u^4
\nonumber\\
&&{} + 80988 t^3 u^5 + 36475 t^2 u^6 + 9328 t u^7 + 
       1079 u^8)
\nonumber\\
&&{} - 2 s (t + u) (91 t^8 + 1293 t^7 u + 6382 t^6 u^2 + 15744 t^5 u^3 + 
  21112 t^4 u^4
\nonumber\\
&&{} + 15744 t^3 u^5 + 6382 t^2 u^6 + 1293 t u^7 + 91 u^8)
\nonumber\\
&&{} - (t + u)^4 (7 t^6 + 228 t^5 u + 1252 t^4 u^2 + 
       1952 t^3 u^3 + 1252 t^2 u^4 + 228 t u^5 + 7 u^6)]
\nonumber\\
&&{} + 2 m_Z^2 [16 s^9 (t + u)^2 + 
     2 s^8 (t + u) (35 t^2 + 82 t u + 35 u^2)
\nonumber\\
&&{} + 
     s^7 (95 t^4 + 627 t^3 u + 1056 t^2 u^2 + 627 t u^3 + 95 u^4)
\nonumber\\
&&{} - 
     s^6 (t + u) (57 t^4 - 721 t^3 u - 1422 t^2 u^2 - 721 t u^3 + 57 u^4)
\nonumber\\
&&{} - s^5 (332 t^6 + 227 t^5 u - 1538 t^4 u^2 - 2874 t^3 u^3 - 
       1538 t^2 u^4 + 227 t u^5 + 332 u^6)
\nonumber\\
&&{} - 
     s^4 (t + u) (424 t^6 + 911 t^5 u + 608 t^4 u^2 + 298 t^3 u^3 + 608 t^2
 u^4 + 
  911 t u^5 + 424 u^6)
\nonumber\\
&&{} - 
     s^3 (253 t^8 + 1415 t^7 u + 3802 t^6 u^2 + 6836 t^5 u^3 + 8376 t^4 u^4
\nonumber\\
&&{} + 
       6836 t^3 u^5 + 3802 t^2 u^6 + 1415 t u^7 + 253 u^8)
\nonumber\\
&&{} - 
     s^2 (t + u) (69 t^8 + 611 t^7 u + 2282 t^6 u^2 + 5134 t^5 u^3 + 
  6750 t^4 u^4 
\nonumber\\
&&{} + 5134 t^3 u^5 + 2282 t^2 u^6 + 611 t u^7 + 69 u^8)
\nonumber\\
&&{} - 
     s (t + u)^2 (6 t^8 + 129 t^7 u + 682 t^6 u^2 + 1796 t^5 u^3 + 
       2452 t^4 u^4
\nonumber\\
&&{} + 1796 t^3 u^5 + 682 t^2 u^6 + 129 t u^7 + 6 u^8)
\nonumber\\
&&{} - 
     t u (t + u)^5 (7 t^4 + 68 t^3 u + 105 t^2 u^2 + 68 t u^3 + 7 u^4)]
\nonumber\\
&&{} + (t + u) (s + t + u)^2 
    [
     s^6 (t + u) (7 t^2 - 4 t u + 7 u^2)
\nonumber\\
&&{} + 
     2 s^5 (14 t^4 + 9 t^3 u - 4 t^2 u^2 + 9 t u^3 + 14 u^4)
\nonumber\\
&&{} + 
     2 s^4 (t + u) (21 t^4 + 3 t^3 u - 4 t^2 u^2 + 3 t u^3 + 21 u^4)
\nonumber\\
&&{} + 2 s^3 (t + u)^2 
      (14 t^4 + 5 t^3 u + 10 t^2 u^2 + 5 t u^3 + 14 u^4)
\nonumber\\
&&{} + s^2 (t + u) (7 t^6 + 38 t^5 u + 64 t^4 u^2 + 116 t^3 u^3 + 64 t^2 u^4 + 
  38 t u^5 + 7 u^6)
\nonumber\\
&&{} + 
     2 s t u (6 t^6 + 25 t^5 u + 69 t^4 u^2 + 94 t^3 u^3 + 69 t^2 u^4 + 
       25 t u^5 + 6 u^6)
\nonumber\\
&&{} + t^2 u^2 (t + u)^3 (7 t^2 + 10 t u + 7 u^2)] ,  
\end{eqnarray}

\begin{equation}
\frac{d\sigma}{dt}\left(\gamma g\to Q\overline{Q}\left[\varsigma^{(8)}\right]
Z\right)
=\frac{9}{32}\,\frac{\alpha_s}{\alpha}\,\frac{d\sigma}{dt}
\left(\gamma\gamma\to Q\overline{Q}\left[\varsigma^{(1)}\right]Z\right),
\qquad\varsigma={}^1\!S_0,{}^3\!S_1,{}^1\!P_1,
\end{equation}

\begin{equation}
\frac{d\sigma}{dt}\left(\gamma g\to Q\overline{Q}
\left[{}^3\!P_J^{(8)}\right]Z\right)
=\frac{9}{32}\,\frac{\alpha_s}{\alpha}\,\sum_{J^\prime=0}^{2}(2J^\prime+1)
\frac{d\sigma}{dt}\left(\gamma\gamma\to Q\overline{Q}
\left[{}^3\!P_{J^\prime}^{(1)}\right]Z\right),
\end{equation}

\begin{equation}
\frac{d\sigma}{dt}\left(gg\to Q\overline{Q}[n]Z\right)
=\frac{9}{512}\,\frac{\alpha_s^2}{\alpha^2}\,
\frac{d\sigma}{dt}\left(\gamma\gamma\to Q\overline{Q}[n]Z\right),\qquad
n={}^1\!S_0^{(1)},{}^3\!S_1^{(1)},{}^1\!P_1^{(1)},{}^3\!P_J^{(1)},
\end{equation}

\begin{eqnarray}
\lefteqn{\frac{d\sigma}{dt}\left(gg\to Q\overline{Q}
\left[{}^1\!S_0^{(8)}\right] Z\right)
=\frac{\pi \alpha_s^2 g^2}{12 M m_Z^2 s^3 (m_Z^2 - s - t)^2 (m_Z^2 - s - u)^2
(2 m_Z^2 - t - u)^2}}
\nonumber\\
&&{} \times \{ 9 v_Q^2 m_Z^2 [9 m_Z^{12} - 3 m_Z^{10} (8 s + 9 (t + u)) 
\nonumber\\
&&{} + 
     m_Z^8 (24 s^2 + 56 s (t + u) + 3 (11 t^2 + 23 t u + 11 u^2))
\nonumber\\
&&{}  - 
     m_Z^6 (10 s^3 + 42 s^2 (t + u) + 2 s (25 t^2 + 54 t u + 25 u^2) + 
       3 (t + u) (7 t^2 + 16 t u + 7 u^2))
\nonumber\\
&&{}  + 
     m_Z^4 (s^4 + 
       12 s^3 (t + u) + s^2 (25 t^2 + 58 t u + 25 u^2) + 
       4 s (t + u) (5 t^2 + 14 t u + 5 u^2)
\nonumber\\
&&{}  + 7 t^4 + 35 t^3 u + 51 t^2 u^2 + 35 t u^3 + 7 u^4)
\nonumber\\
&&{}  - m_Z^2 (s^4 (t + u) + 2 s^3 (2 t^2 + 5 t u + 2 u^2) + 
       5 s^2 (t + u) (t^2 + 4 t u + u^2)
\nonumber\\
&&{}  + 
       s (t + u)^2 (3 t^2 + 16 t u + 3 u^2) + (t + u) (t^2 + t u + u^2) (t^2 +
 7 t u + u^2))
\nonumber\\
&&{}  + 
     t u (s^4 + 2 s^3 (t + u) + 3 s^2 (t + u)^2 + 2 s (t + u)^3 + 
       (t^2 + t u + u^2)^2)]
\nonumber\\
&&{}  + 
  5 a_Q^2 s (2 m_Z^2 - s - t - u)^2 
    [m_Z^8 - 2 m_Z^6 (t + u) - 
     m_Z^4 (s^2 - t^2 - 4 t u - u^2)
\nonumber\\
&&{}  - 2 m_Z^2 t u (t + u) + t^2 u^2] \} ,
\end{eqnarray}

\begin{eqnarray}
\lefteqn{\frac{d\sigma}{dt}\left(gg\to Q\overline{Q}
\left[{}^3\!S_1^{(8)}\right]Z\right)
=\frac{-\pi \alpha_s^2 g^2}{36 M m_Z^2 s^3 (m_Z^2 - s - t)^2 (m_Z^2 - s - u)^2
(2 m_Z^2 - t - u)^2}}
\nonumber\\
&&{} \times \{5 v_Q^2 m_Z^2 s [m_Z^{10} - 4 m_Z^8 (3 s + t + u) + 
     m_Z^6 (22 s^2 + 26 s (t + u) + 5 t^2 + 12 t u + 5 u^2)
\nonumber\\
&&{} - 2 m_Z^4 (5 s^3 + 14 s^2 (t + u) + s (8 t^2 + 23 t u + 8 u^2) + (t + u)
 (t^2 + 5 t u + u^2))
\nonumber\\
&&{}  - 
     m_Z^2 (s^4 - 4 s^3 (t + u) - 
       s^2 (9 t^2 + 26 t u + 9 u^2) - 2 s (t + u) (t^2 + 10 t u + u^2)
\nonumber\\
&&{}  - t u (4 t^2 + 9 t u + 4 u^2))
\nonumber\\
&&{}  - 
     2 (s^3 (t^2 + t u + u^2) + s^2 (t + u)^3 + 
       s t u (t^2 + 3 t u + u^2) + t^2 u^2 (t + u))]
\nonumber\\
&&{}  
    - 9 a_Q^2 [9 m_Z^{12} (s + t + u) - m_Z^{10} (18 s^2 + 49 s (t + u) + 
       27 (t + u)^2)
\nonumber\\
&&{}  + m_Z^8 (7 s^3 + 57 s^2 (t + u) + 
       s (83 t^2 + 169 t u + 83 u^2) + 3 (t + u) (11 t^2 + 23 t u + 11 u^2))
\nonumber\\
&&{}  + m_Z^6 (2 s^4 - 13 s^3 (t + u) - 
       2 s^2 (31 t^2 + 63 t u + 31 u^2) - 5 s (t + u) (13 t^2 + 28 t u + 13 u^2)
\nonumber\\
&&{}  - 
       3 (t + u)^2 (7 t^2 + 16 t u + 7 u^2))
\nonumber\\
&&{}  - m_Z^4 (2 s^4 (t + u) - s^3 (8 t^2 + 19 t u + 8 u^2) - 
       5 s^2 (t + u) (3 t + 2 u) (2 t + 3 u)
\nonumber\\
&&{}  - 
       5 s (5 t^4 + 23 t^3 u + 35 t^2 u^2 + 23 t u^3 + 5 u^4)
\nonumber\\
&&{}  - (t + u) (7 t^4 + 35 t^3 u + 51 t^2 u^2 + 35 t u^3 + 7 u^4))
\nonumber\\
&&{}  + 
     m_Z^2 (s^4 (t^2 + u^2) - 
       2 s^3 (t + u) (t^2 + 3 t u + u^2) - s^2 (t + u)^2 (6 t^2 + 17 t u + 6
 u^2)
\nonumber\\
&&{}  - 
       s (t + u) (4 t^4 + 26 t^3 u + 37 t^2 u^2 + 26 t u^3 + 4 u^4)
\nonumber\\
&&{}  - 
       (t + u)^2 (t^4 + 8 t^3 u + 9 t^2 u^2 + 8 t u^3 + u^4))
\nonumber\\
&&{}  + t u (s + t + u) 
      (s (t + u) + t^2 + t u + u^2)^2]\} ,
\end{eqnarray}

\begin{eqnarray}
\lefteqn{\frac{d\sigma}{dt}\left(gg\to Q\overline{Q}
\left[{}^1\!P_1^{(8)}\right]Z\right)
=\frac{-\pi \alpha_s^2 g^2}{9 M^3 m_Z^2 s^3 (m_Z^2 - s - t)^3
(m_Z^2 - s - u)^3  (2 m_Z^2 - t - u)^4}}
\nonumber\\
&&{} \times \{   10 v_Q^2 m_Z^2 s [ 48 m_Z^{18}
- 16 m_Z^{16} (11 s + 15 (t + u)) 
\nonumber\\
&&{} + 2 m_Z^{14} (141 s^2 + 374 s (t + u) + 4 (65 t^2 + 134 t u + 65 u^2))
\nonumber\\
&&{} - 4 m_Z^{12} (65 s^3 + 255 s^2 (t + u) + s (341 t^2 + 696 t u + 341 u^2)
\nonumber\\
&&{} + 32 (t + u) (5 t^2 + 11 t u + 5 u^2))
\nonumber\\
&&{} + m_Z^{10} (138 s^4 + 798 s^3 (t + u) + 
        10 s^2 (158 t^2 + 315 t u + 158 u^2)
\nonumber\\
&&{}  + 
        s (t + u) (1399 t^2 + 2958 t u + 1399 u^2)
\nonumber\\
&&{}  + 491 t^4 + 2188 t^3 u + 
        3378 t^2 u^2 + 2188 t u^3 + 491 u^4)
\nonumber\\
&&{}  - 
      m_Z^8 (24 s^5 + 356 s^4 (t + u) + 
        20 s^3 (53 t^2 + 101 t u + 53 u^2)
\nonumber\\
&&{}  + 10 s^2 (t + u) (138 t^2 + 265 t u + 138 u^2) + s (t + u)^2 
         (887 t^2 + 1956 t u + 887 u^2)
\nonumber\\
&&{}  + (t + u) (241 t^4 + 1188 t^3 u + 1846 t^2 u^2 + 1188 t u^3 + 241 u^4))
\nonumber\\
&&{}  - 
      m_Z^6 (18 s^6 - 60 s^5 (t + u) - s^4 (417 t^2 + 734 t u + 417 u^2)
\nonumber\\
&&{} 
      - 
        s^3 (t + u) (801 t^2 + 1298 t u + 801 u^2)
\nonumber\\
&&{}  - 
        s^2 (749 t^4 + 2772 t^3 u + 4068 t^2 u^2 + 2772 t u^3 + 749 u^4)
\nonumber\\
&&{}  - 
        s (t + u) (359 t^4 + 1534 t^3 u + 2362 t^2 u^2 + 1534 t u^3 + 359 u^4)
\nonumber\\
&&{}  - 
        2 (t + u)^2 (37 t^4 + 217 t^3 u + 332 t^2 u^2 + 217 t u^3 + 37 u^4))
\nonumber\\
&&{}  + m_Z^4 (12 s^7 + 16 s^6 (t + u) - 
        s^5 (83 t^2 + 122 t u + 83 u^2)
\nonumber\\
&&{}  - s^4 (t + u) (275 t^2 + 356 t u + 275 u^2)
\nonumber\\
&&{}  - s^3 (367 t^4 + 1153 t^3 u + 
          1580 t^2 u^2 + 1153 t u^3 + 367 u^4)
\nonumber\\
&&{}  - 
        s^2 (t + u) (258 t^4 + 843 t^3 u + 1214 t^2 u^2 + 843 t u^3 + 258 u^4)
\nonumber\\
&&{}  - 4 s (23 t^6 + 143 t^5 u + 371 t^4 u^2 + 
          500 t^3 u^3 + 371 t^2 u^4 + 143 t u^5 + 23 u^6)
\nonumber\\
&&{}  - (t + u)^3 (13 t^4 + 104 t^3 u + 
          150 t^2 u^2 + 104 t u^3 + 13 u^4))
\nonumber\\
&&{}  - m_Z^2 (2 s^8 + 6 s^7 (t + u) - 
        s^6 (7 t^2 + 4 t u + 7 u^2) - 52 s^5 (t + u) (t^2 + t u + u^2)
\nonumber\\
&&{}  - 
        s^4 (97 t^4 + 269 t^3 u + 346 t^2 u^2 + 269 t u^3 + 97 u^4)
\nonumber\\
&&{}  - 
        2 s^3 (t + u) (48 t^4 + 122 t^3 u + 155 t^2 u^2 + 122 t u^3 + 48 u^4)
\nonumber\\
&&{}  - s^2 (53 t^6 + 244 t^5 u + 531 t^4 u^2 + 674 t^3 u^3 + 
          531 t^2 u^4 + 244 t u^5 + 53 u^6)
\nonumber\\
&&{}  - 
        s (t + u) (14 t^6 + 82 t^5 u + 217 t^4 u^2 + 290 t^3 u^3 + 217 t^2 u^4
 + 
  82 t u^5 + 14 u^6)
\nonumber\\
&&{}  - (t + u)^4 (t^4 + 15 t^3 u + 19 t^2 u^2 + 15 t u^3 + u^4))
\nonumber\\
&&{}  - 
      (t + u) (s^7 (t + u) + 
        s^6 (5 t^2 + 6 t u + 5 u^2) + s^5 (t + u) (11 t^2 + 8 t u + 11 u^2)
\nonumber\\
&&{}  + 2 s^4 (7 t^4 + 16 t^3 u + 19 t^2 u^2 + 16 t u^3 + 
          7 u^4)
\nonumber\\
&&{}  + s^3 (t + u) (11 t^4 + 21 t^3 u + 25 t^2 u^2 + 21 t u^3 + 11 u^4)
\nonumber\\
&&{}  + s^2 (5 t^6 + 19 t^5 u + 38 t^4 u^2 + 
          46 t^3 u^3 + 38 t^2 u^4 + 19 t u^5 + 5 u^6)
\nonumber\\
&&{}  + 
        s (t + u) (t^6 + 5 t^5 u + 14 t^4 u^2 + 18 t^3 u^3 + 14 t^2 u^4 + 5 t
 u^5 + 
  u^6)
\nonumber\\
&&{}  + t u (t + u)^4 (t^2 + t u + u^2)) ]
\nonumber\\
&&{} + 9
  a_Q^2 [ 36 m_Z^{20} (s - t - u) - 
      4 m_Z^{18} (17 s^2 + 8 s (t + u) - 45 (t + u)^2)
\nonumber\\
&&{}  + 
      m_Z^{16} (28 s^3 + 12 s^2 (t + u) - s (257 t^2 + 466 t u + 257 u^2)
\nonumber\\
&&{}  - 
        3 (t + u) (131 t^2 + 278 t u + 131 u^2))
\nonumber\\
&&{}  - 
      4 m_Z^{14} (2 s^4 - 40 s^3 (t + u) - 6 s^2 (22 t^2 + 41 t u + 22
      u^2)
\nonumber\\
&&{}  - 
        4 s (t + u) (47 t^2 + 88 t u + 47 u^2) - 
        3 (t + u)^2 (41 t^2 + 98 t u + 41 u^2))
\nonumber\\
&&{}  + 
      m_Z^{12} (28 s^5 - 88 s^4 (t + u) - s^3 (695 t^2 + 1334 t u + 695 u^2)
\nonumber\\
&&{}  - 
        s^2 (t + u) (1201 t^2 + 2218 t u + 1201 u^2)
\nonumber\\
&&{}  - 
        2 s (485 t^4 + 1938 t^3 u + 2910 t^2 u^2 + 1938 t u^3 + 485 u^4)
\nonumber\\
&&{}  - 
        4 (t + u) (97 t^4 + 473 t^3 u + 750 t^2 u^2 + 473 t u^3 + 97 u^4))
\nonumber\\
&&{}  - m_Z^{10} (20 s^6 + 44 s^5 (t + u) - 
        26 s^4 (11 t^2 + 20 t u + 11 u^2)
\nonumber\\
&&{}  - 8 s^3 (t + u) (132 t^2 + 241 t u + 132 u^2)
\nonumber\\
&&{}  - s^2 (1277 t^4 + 4942 t^3 u + 
          7386 t^2 u^2 + 4942 t u^3 + 1277 u^4)
\nonumber\\
&&{}  - 
        2 s (t + u) (363 t^4 + 1544 t^3 u + 2346 t^2 u^2 + 1544 t u^3 + 363 u^4)
\nonumber\\
&&{}  - 6 (t + u)^2 (33 t^4 + 190 t^3 u + 
          310 t^2 u^2 + 190 t u^3 + 33 u^4))
\nonumber\\
&&{}  + m_Z^8 (4 s^7 + 32 s^6 (t + u) + 
        s^5 (7 t^2 + 30 t u + 7 u^2)
\nonumber\\
&&{}  - s^4 (t + u) (351 t^2 + 566 t u + 351 u^2)
\nonumber\\
&&{}  - 2 s^3 (421 t^4 + 1577 t^3 u + 2354 t^2 u^2 + 
          1577 t u^3 + 421 u^4)
\nonumber\\
&&{}  - s^2 (t + u) (779 t^4 + 3112 t^3 u + 4734 t^2 u^2 + 3112 t u^3 + 779 u^4)
\nonumber\\
&&{}  - 
        2 s (168 t^6 + 1138 t^5 u + 2987 t^4 u^2 + 4042 t^3 u^3 + 
          2987 t^2 u^4 + 1138 t u^5 + 168 u^6)
\nonumber\\
&&{}  - 
        2 (t + u) (32 t^6 + 303 t^5 u + 915 t^4 u^2 + 1280 t^3 u^3 + 915 t^2
 u^4 + 
  303 t u^5 + 32 u^6))
\nonumber\\
&&{}  - 
      m_Z^6 (4 s^7 (t + u) + 2 s^6 (9 t^2 + 16 t u + 9 u^2) - 
        4 s^5 (t + u) (5 t^2 + 6 t u + 5 u^2)
\nonumber\\
&&{}  - 
        2 s^4 (107 t^4 + 374 t^3 u + 550 t^2 u^2 + 374 t u^3 + 107 u^4)
\nonumber\\
&&{}  - 
        3 s^3 (t + u) (127 t^4 + 478 t^3 u + 742 t^2 u^2 + 478 t u^3 + 127 u^4)
\nonumber\\
&&{}  - 
        2 s^2 (141 t^6 + 897 t^5 u + 2309 t^4 u^2 + 3134 t^3 u^3 + 
          2309 t^2 u^4 + 897 t u^5 + 141 u^6)
\nonumber\\
&&{}  - 
        s (t + u) (95 t^6 + 746 t^5 u + 2017 t^4 u^2 + 2764 t^3 u^3 + 2017 t^2
 u^4 + 
  746 t u^5 + 95 u^6)
\nonumber\\
&&{}  - 4 (t + u)^2 (3 t^6 + 40 t^5 u + 
          134 t^4 u^2 + 186 t^3 u^3 + 134 t^2 u^4 + 40 t u^5 + 3 u^6))
\nonumber\\
&&{}  + 
      m_Z^4 (s^7 (t + u)^2 + s^6 (t + u) (5 t^2 + 2 t u + 5 u^2)
\nonumber\\
&&{}  - 
        2 s^5 (5 t^4 + 24 t^3 u + 36 t^2 u^2 + 24 t u^3 + 5 u^4)
\nonumber\\
&&{}  - 
        s^4 (t + u) (65 t^4 + 240 t^3 u + 374 t^2 u^2 + 240 t u^3 + 65 u^4)
\nonumber\\
&&{}  - s^3 (94 t^6 + 577 t^5 u + 1489 t^4 u^2 + 2060 t^3 u^3 + 
          1489 t^2 u^4 + 577 t u^5 + 94 u^6)
\nonumber\\
&&{}  - 
        s^2 (t + u) (57 t^6 + 424 t^5 u + 1100 t^4 u^2 + 1578 t^3 u^3 + 
  1100 t^2 u^4 + 424 t u^5 + 57 u^6)
\nonumber\\
&&{}  - 
        s (15 t^8 + 186 t^7 u + 758 t^6 u^2 + 1626 t^5 u^3 + 2074 t^4 u^4
\nonumber\\
&&{}  + 
          1626 t^3 u^5 + 758 t^2 u^6 + 186 t u^7 + 15 u^8)
\nonumber\\
&&{}  - (t + u) (t^8 + 28 t^7 u + 160 t^6 u^2 + 376 t^5 u^3 + 490 t^4 u^4
\nonumber\\
&&{}  + 
   376 t^3 u^5 + 160 t^2 u^6 + 28 t u^7 + u^8))
\nonumber\\
&&{}  - 
      m_Z^2 (s^6 (t - u)^2 (t + u)^2
- s^5 (t + u) (t^4 + 12 t^3 u + 18 t^2 u^2 + 12 t u^3 + u^4)
\nonumber\\
&&{}  - 2 s^4 (4 t^6 + 29 t^5 u + 
          77 t^4 u^2 + 110 t^3 u^3 + 77 t^2 u^4 + 29 t u^5 + 4 u^6)
\nonumber\\
&&{}  - 
        2 s^3 (t + u) (5 t^6 + 42 t^5 u + 106 t^4 u^2 + 170 t^3 u^3 + 106 t^2
 u^4 + 
  42 t u^5 + 5 u^6)
\nonumber\\
&&{}  - s^2 (5 t^8 + 69 t^7 u + 
          261 t^6 u^2 + 577 t^5 u^3 + 756 t^4 u^4 + 577 t^3 u^5 + 
          261 t^2 u^6 + 69 t u^7 + 5 u^8)
\nonumber\\
&&{}  - s (t + u)^3 (t^6 + 19 t^5 u + 53 t^4 u^2 + 78 t^3 u^3 + 
          53 t^2 u^4 + 19 t u^5 + u^6)
\nonumber\\
&&{}  - 2 t u (t + u)^2 
         (t^6 + 9 t^5 u + 21 t^4 u^2 + 28 t^3 u^3 + 21 t^2 u^4 + 9 t u^5 + 
          u^6))
\nonumber\\
&&{}  - t u (t + u) (s^5 (t + u) (t^2 + t u + u^2) + 
        s^4 (4 t^4 + 9 t^3 u + 18 t^2 u^2 + 9 t u^3 + 4 u^4)
\nonumber\\
&&{}  + 
        2 s^3 (t + u) (3 t^4 + 5 t^3 u + 14 t^2 u^2 + 5 t u^3 + 3 u^4)
\nonumber\\
&&{}  + 2 s^2 (t + u)^2 
         (2 t^4 + 3 t^3 u + 9 t^2 u^2 + 3 t u^3 + 2 u^4)
\nonumber\\
&&{}  + 
        s (t + u) (t^3 + 3 t^2 u + 2 t u^2 + u^3) (t^3 + 2 t^2 u + 3 t u^2 +
 u^3)
\nonumber\\
&&{}  + 
        t u (t + u)^2 (t^2 + t u + u^2)^2) ]
 \} ,
\end{eqnarray}

\begin{eqnarray}
\lefteqn{\frac{d\sigma}{dt}\left(gg\to Q\overline{Q}
\left[{}^3\!P_J^{(8)}\right]Z\right)
=\frac{\pi \alpha_s^2 g^2}{3 M^3 m_Z^2 s^3 (m_Z^2 - s - t)^3 (m_Z^2 - s - u)^3
 (2 m_Z^2 - t - u)^4}}
\nonumber\\
&&{} \times \{9 v_Q^2 m_Z^2 [ 72 m_Z^{20} - 12 m_Z^{18} (31 s + 33 (t + u))
\nonumber\\
&&{}  +  m_Z^{16}
    (760 s^2 + 1788 s (t
    + u) + 975 t^2 + 2046 t u + 975 u^2)
\nonumber\\
&&{}  -  m_Z^{14} (804 s^3 + 3144 s^2 (t + u) + s
    (3785 t^2 + 7854 t u +  3785 u^2)
\nonumber\\
&&{}  + 108 (t + u) (13 t^2 + 30 t u + 13 u^2))
\nonumber\\
&&{}  +  m_Z^{12} (520 s^4 +
    2868 s^3 (t + u) + s^2 (5661 t^2 + 11578 t u +  5661
    u^2)
\nonumber\\
&&{}  + 4 s (t + u) (1151 t^2 + 2569 t u + 1151 u^2)
\nonumber\\
&&{}  +  4 (323 t^4 + 1513 t^3 u + 2376 t^2 u^2
    + 1513 t u^3 + 323 u^4))
\nonumber\\
&&{}  -  m_Z^{10} (236 s^5 + 1644
    s^4 (t + u) + s^3 (4431 t^2 + 8902 t u +  4431 u^2) 
\nonumber\\
&&{}
    +  2 s^2 (t + u) (2885 t^2 + 6188 t u + 2885 u^2)
\nonumber\\
&&{} +  s (3514 t^4 + 15863 t^3 u + 24534 t^2 u^2 +
    15863 t u^3 + 3514 u^4)
\nonumber\\
&&{}  +  2 (t + u) (389 t^4 + 2086 t^3 u + 3366 t^2 u^2 + 2086 t u^3 + 389 u^4))
\nonumber\\
&&{}  + m_Z^8 (56 s^6 +  628 s^5 (t + u) + s^4 (2199 t^2 + 4382
    t u + 2199 u^2)
\nonumber\\
&&{} +  20 s^3 (t + u) (190 t^2 + 393 t u + 190 u^2)
\nonumber\\
&&{}  +  s^2 (3606 t^4 + 15715 t^3 u +
    23922 t^2 u^2 + 15715 t u^3 +  3606 u^4)
\nonumber\\
&&{}  + s (t + u) (1720 t^4 + 8657 t^3 u + 13402 t^2 u^2 + 8657 t u^3 + 1720
 u^4)
\nonumber\\
&&{} + 2 (150 t^6 + 1301 t^5 u + 3823 t^4 u^2 + 5328 t^3 u^3 + 3823 t^2 u^4 + 
  1301 t u^5 + 150 u^6))
\nonumber\\
&&{}  + m_Z^6 (4 s^7 - 96 s^6 (t +
    u) -  3 s^5 (219 t^2 + 442 t u + 219 u^2)
\nonumber\\
&&{} - 4 s^4 (t + u) (387 t^2 + 800 t u + 387 u^2)
\nonumber\\
&&{} - 4
    s^3 (479 t^4 + 2045 t^3 u +  3095 t^2 u^2 + 2045 t
    u^3 + 479 u^4)
\nonumber\\
&&{}  -  2 s^2 (t + u) (695 t^4 + 3346 t^3 u + 5022 t^2 u^2 + 3346 t u^3 + 695
 u^4)
\nonumber\\
&&{} - s
    (526 t^6 + 4237 t^5 u + 11717 t^4 u^2 +  15936 t^3
    u^3 + 11717 t^2 u^4 + 4237 t u^5 + 526 u^6)
\nonumber\\
&&{}   -  4 (t + u) (17 t^6 + 198 t^5 u + 636 t^4 u^2 + 890 t^3 u^3 + 636 t^2
 u^4 + 
  198 t u^5 + 17 u^6) )
\nonumber\\
&&{} - 
    m_Z^4 (4 s^7 (t + u) -  69 s^6 (t +
    u)^2 - 20 s^5 (t + u) (17 t^2 + 36 t u + 17 u^2)
\nonumber\\
&&{}  -  2 s^4 (297 t^4 + 1276 t^3 u + 1954 t^2 u^2 + 1276 t u^3 + 297 u^4)
\nonumber\\
&&{}  -  2 s^3 (t + u) (276 t^4 + 1311 t^3 u + 1996 t^2 u^2 + 1311 t u^3 + 276
 u^4)
\nonumber\\
&&{} 
    - s^2 (312 t^6 + 2433 t^5 u + 6563 t^4 u^2 +  8836
    t^3 u^3 + 6563 t^2 u^4 + 2433 t u^5 + 312 u^6)
\nonumber\\
&&{}  -  s (t + u) (92 t^6 + 947 t^5 u + 2734 t^4 u^2 + 3626 t^3 u^3 + 2734
 t^2 u^4 + 
  947 t u^5 + 92 u^6)
\nonumber\\
&&{}  - 7 t^8 - 164 t^7 u - 876 t^6 u^2 - 2048 t^5
    u^3 - 2666 t^4 u^4
\nonumber\\
&&{} -  2048 t^3 u^5 - 876 t^2 u^6 -
    164 t u^7 - 7 u^8)
\nonumber\\
&&{}  +  m_Z^2 (s^7 (t + u)^2 - 2 s^6 (t + u) (11 t^2 + 28 t u + 11 u^2)
\nonumber\\
&&{}  -  s^5 (80
    t^4 + 387 t^3 u + 590 t^2 u^2 + 387 t u^3 + 80 u^4)
\nonumber\\
&&{} 
    -  4 s^4 (t + u) (26 t^4 + 143 t^3 u + 215 t^2 u^2 + 143 t u^3 + 26 u^4)
\nonumber\\
&&{}  -  s^3 (72 t^6 + 635 t^5 u +
    1741 t^4 u^2 + 2396 t^3 u^3 +  1741 t^2 u^4 + 635 t
    u^5 + 72 u^6)
\nonumber\\
&&{}  - 2 s^2 (t + u) (16 t^6 + 180 t^5 u + 495 t^4 u^2 + 672 t^3 u^3 + 495
 t^2 u^4 + 
  180 t u^5 + 16 u^6)
\nonumber\\
&&{}  - s (t + u)^2 (7 t^6 +
    123 t^5 u + 376 t^4 u^2 +  465 t^3 u^3 + 376 t^2 u^4
    + 123 t u^5 + 7 u^6)
\nonumber\\
&&{}  -  2 t u (t + u) (t^2 + t u + u^2) (7 t^4 + 48 t^3 u + 76 t^2 u^2 + 48 t
 u^3 + 
  7 u^4))
\nonumber\\
&&{}  +  (t + u) (s^6 (t + u) (2 t^2 + 11 t u + 2 u^2) +  s^5 (4 t^4 + 47 t^3
 u + 70 t^2 u^2
    + 47 t u^3 + 4 u^4)
\nonumber\\
&&{}  +  s^4 (t + u) (2 t^4 + 61 t^3 u + 75 t^2 u^2 + 61 t u^3 + 2 u^4)
\nonumber\\
&&{} +  s^3 t u (47 t^4 + 132 t^3 u +
    190 t^2 u^2  + 132 t u^3 + 47 u^4)
\nonumber\\
&&{} + s^2 t u (t + u) (25 t^4 + 63 t^3 u + 93 t^2 u^2 + 63 t u^3 + 25 u^4)
\nonumber\\
&&{}  + s t u (7 t^6 +
    38 t^5 u + 78 t^4 u^2 +  98 t^3 u^3 + 78 t^2 u^4 +
    38 t u^5 + 7 u^6)
\nonumber\\
&&{}  + 7 t^2 u^2 (t +
    u) (t^2 + t u + u^2)^2) ]
\nonumber\\
&&{} - 10
    a_Q^2 s [ 36 m_Z^{20} - 8 m_Z^{18} (16 s + 17 (t + u))
\nonumber\\
&&{}  + 
    m_Z^{16} (212 s^2 + 320
    s (t + u) + 153 t^2 + 382 t u + 153 u^2)
\nonumber\\
&&{}  -  m_Z^{14} (318 s^3 +  346 s^2 (t + u) - 2 s (11 t^2 - 114 t u + 11 u^2)
\nonumber\\
&&{} 
- (t + u) (67 t^2 - 150 t u + 67 u^2))
\nonumber\\
&&{}  +  m_Z^{12} (440 s^4 +
    636 s^3 (t + u) - s^2 (373 t^2 + 362 t u + 373 u^2)
\nonumber\\
&&{} 
    -  s (t + u) (915 t^2 + 1034 t u + 915 u^2)
\nonumber\\
&&{}  - 337 t^4 -
    949 t^3 u - 1180 t^2 u^2 - 949 t u^3 -  337 u^4)
\nonumber\\
&&{} 
    -  m_Z^{10} (388 s^5 + 982 s^4 (t + u) + 2 s^3 (128
    t^2 + 417 t u +  128 u^2)
\nonumber\\
&&{}  - s^2 (t + u) (1351 t^2 + 1864 t u + 1351 u^2)
\nonumber\\
&&{} -  2 s (687 t^4 + 2335
    t^3 u + 3216 t^2 u^2 + 2335 t u^3 + 687 u^4)
\nonumber\\
&&{}  -  2 (t + u) (187 t^4 + 636 t^3 u + 818 t^2 u^2 + 636 t u^3 + 187 u^4))
\nonumber\\
&&{}  + m_Z^8 (188 s^6  +  718 s^5 (t + u) + 2
    s^4 (421 t^2 + 946 t u + 421 u^2)
\nonumber\\
&&{}  -  2 s^3 (t + u) (151 t^2 + 45 t u + 151 u^2)
\nonumber\\
&&{}  -  2 s^2 (703 t^4
    + 2496 t^3 u + 3508 t^2 u^2 + 2496 t u^3 + 703 u^4)
\nonumber\\
&&{} 
    -  s (t + u) (1003 t^4 + 3708 t^3 u + 5034 t^2 u^2 + 3708 t u^3 + 1003 u^4)
\nonumber\\
&&{}  - 217 t^6 -
    1323 t^5 u - 3116 t^4 u^2 -  4016 t^3 u^3 - 3116 t^2
    u^4 - 1323 t u^5 - 217 u^6)
\nonumber\\
&&{}  - m_Z^6 (46 s^7 +  240 s^6
    (t + u) + 5 s^5 (101 t^2 + 220 t u + 101 u^2)
\nonumber\\
&&{}  +  s^4 (t + u) (339 t^2 + 916 t u + 339 u^2)
\nonumber\\
&&{} 
 -  2
    s^3 (196 t^4 + 627 t^3 u + 833 t^2 u^2 + 627 t u^3 +
    196 u^4)
\nonumber\\
&&{}  -  s^2 (t + u) (771 t^4 + 2854 t^3 u + 3968 t^2 u^2 + 2854 t u^3 + 771
 u^4)
\nonumber\\
&&{}  - s (421 t^6
    + 2566 t^5 u + 6173 t^4 u^2 +  8040 t^3 u^3 + 6173
    t^2 u^4 + 2566 t u^5 + 421 u^6)
\nonumber\\
&&{}  - (t + u) (72 t^6 + 524 t^5 u + 1295 t^4 u^2 + 1674 t^3 u^3 + 1295 t^2
 u^4 + 
   524 t u^5 + 72 u^6))
\nonumber\\
&&{}  +  m_Z^4 (4 s^8 + 30 s^7 (t + u) + 2 s^6 (53
    t^2 + 114 t u + 53 u^2)
\nonumber\\
&&{}  + s^5 (t + u) (163 t^2 + 388 t u + 163 u^2)  +  s^4 (29 t^4 + 237 t^3 u +
    424 t^2 u^2 + 237 t u^3 + 29 u^4)
\nonumber\\
&&{}  -  2 s^3 (t + u)^3 (108 t^2
    + 127 t u + 108 u^2)
\nonumber\\
&&{}  - s^2 (254 t^6 + 1423
    t^5 u +  3385 t^4 u^2 + 4416 t^3 u^3 + 3385 t^2 u^4
    + 1423 t u^5 +  254 u^6)
\nonumber\\
&&{}  - s (t + u) (105 t^6 + 682 t^5 u + 1694 t^4 u^2 + 2206 t^3 u^3 + 1694
 t^2 u^4 + 
  682 t u^5 + 105 u^6)
\nonumber\\
&&{}  -  (t + u)^2 (13
    t^6 + 126 t^5 u + 331 t^4 u^2 + 427 t^3 u^3 +  331
    t^2 u^4 + 126 t u^5 + 13 u^6))
\nonumber\\
&&{}  -  m_Z^2 (5 s^7 (t
    + u)^2 + s^6 (t + u) (17 t^2 + 38 t u + 17 u^2)
\nonumber\\
&&{} 
    +  s^5 (11 t^4 + 67 t^3 u + 110 t^2 u^2 + 67 t u^3 +
    11 u^4)
\nonumber\\
&&{}  -  s^4 (t + u) (3 t^2 + 5 t u + 3 u^2) (11 t^2 + 2 t u + 11 u^2)
\nonumber\\
&&{}  -  s^3 (69 t^6 + 310 t^5 u + 677 t^4 u^2 +
    866 t^3 u^3 + 677 t^2 u^4 +  310 t u^5 + 69 u^6)
\nonumber\\
&&{}  -
    s^2 (t + u) (51 t^6 + 252 t^5 u + 598 t^4 u^2 + 776 t^3 u^3 + 598 t^2 u^4 + 
  252 t u^5 + 51 u^6)
\nonumber\\
&&{}  - s (t + u)^2 (15 t^6 + 99 t^5 u + 
    259 t^4 u^2 + 336 t^3 u^3 + 259 t^2 u^4 + 99 t u^5 +
    15 u^6)
\nonumber\\
&&{}  - (t + u)^3 (t^6 +
    17 t^5 u + 48 t^4 u^2 + 62 t^3 u^3 +  48 t^2 u^4 +
    17 t u^5 + u^6))
\nonumber\\
&&{}  - (t + u) (s + t
    + u)^2  (s^4 (t + u) (t^2 + t u + u^2) + s^3 (3 t^4 + 6 t^3 u +
    8 t^2 u^2 + 6 t u^3 + 3 u^4) 
\nonumber\\
&&{} +   s^2 (t + u) (3 t^4 + 4 t^3 u + 9 t^2 u^2 + 4 t u^3 + 3 u^4)
\nonumber\\
&&{} +  s
    (t^6 + 4 t^5 u + 11 t^4 u^2 + 14 t^3 u^3 + 11 t^2
    u^4 + 4 t u^5 +  u^6)
\nonumber\\
&&{} + t u (t + u)^3 (t^2 + t u + u^2)) ]  \} ,
\end{eqnarray}

\begin{equation}
\frac{d\sigma}{dt}\left(q\overline{q}\to Q\overline{Q}
\left[{}^1\!S_0^{(8)}\right]Z\right)
=\frac{4 \pi \alpha_s^2 g^2 v_Q^2}{9 M s^3 (2 m_Z^2 - t - u)^2}
[2 m_Z^4 - 2 m_Z^2 (s + t + u) + t^2 + u^2], \qquad
\end{equation}

\begin{eqnarray}
\lefteqn{\frac{d\sigma}{dt}\left(q\overline{q}\to Q\overline{Q}
\left[{}^3\!S_1^{(8)}\right]Z\right)
=\frac{-2 \pi \alpha_s^2 g^2}{27 M^5 m_Z^2 s^3 t^2 u^2 (2 m_Z^2 - t - u)^2}}
\nonumber\\
&&{} \times \{(v_q^2 + a_q^2) m_Z^2 s (2 m_Z^2 - t - u)^2  [ m_Z^6 (t^2 +
   u^2) -  2 m_Z^4 (s + t + u) (t^2 + u^2) 
\nonumber\\
&&{} +  m_Z^2 (s^2 (t + u)^2 + 2 s (t + u) (t^2 + t u + u^2)  + (t^2 + u^2)
 (t^2 + 3 t u + u^2) )
\nonumber\\
&&{}  - t u (2
   s^3 + 4 s^2 (t + u) +
   s (3 t^2 + 4 t u + 3 u^2) + (t + u) (t^2 + u^2))]
\nonumber\\
&&{}  -4 a_q a_Q M^2 m_Z^2 s t u (2
   m_Z^2 - t - u) (t + u)[2 m_Z^4 - 3 m_Z^2 (s +
   t + u) + (s + t + u)^2]
\nonumber\\
&&{}  + 
 2 a_Q^2 M^4 t^2 u^2 [ 2
   m_Z^4 (3 s - t - u)  - 
   m_Z^2 (6 s^2 + 4 s (t + u) - 2 (t + u)^2)
\nonumber\\
&&{}  - (s + t + u) (t^2 + u^2)] \} ,
\end{eqnarray}

\begin{eqnarray}
\lefteqn{\frac{d\sigma}{dt}\left(q\overline{q}\to Q\overline{Q}
\left[{}^1\!P_1^{(8)}\right]Z\right)
=\frac{-16 \pi \alpha_s^2 g^2 a_Q^2}{27 M^3 m_Z^2 s^3 (2 m_Z^2 - t - u)^4}}
\nonumber\\
&&{} \times \{ 8 m_Z^8 (s - t - u) - 16 m_Z^6 [s^2 - (t + u)^2]
\nonumber\\
&&{}  + 
   2 m_Z^4 [4 s^3 + 4 s^2 (t + u) - 
     s (3 t^2 + 10 t u + 3 u^2) - (t + u) (7 t^2 + 10 t u + 7 u^2)]
\nonumber\\
&&{}  - 
   2 m_Z^2 [s^2 (t - u)^2 - 2 s (t + u)^3 - 
     (t + u)^2 (3 t^2 + 2 t u + 3 u^2)]
\nonumber\\
&&{}  - (s + t + u)  (t + u)^2
    (t^2 + u^2)\} ,
\end{eqnarray}

\begin{eqnarray}
\lefteqn{\frac{d\sigma}{dt}\left(q\overline{q}\to Q\overline{Q}
\left[{}^3\!P_J^{(8)}\right]Z\right)
=
\frac{16 \pi \alpha_s^2 g^2 v_Q^2}{9 M^3 s^3 (2 m_Z^2 - t - u)^4}}
\nonumber\\
&&{} \times \{ 16 m_Z^8 - 8
   m_Z^6 [8 s + 5 (t + u)] + 2 m_Z^4 [16 s^2 + 48 s
   (t + u) + 23 t^2 + 38 t u + 23 u^2]
\nonumber\\
&&{}  +  2 m_Z^2 [8 s^3 - 4 s^2
   (t + u) -  23 s (t + u)^2 - (t + u) (13 t^2 + 14 t u + 13 u^2)]
\nonumber\\
&&{}  + (t + u) [8 s^2 (t + u) + 4 s (3 t^2 + 4 t u + 3
   u^2) +  7 (t + u) (t^2 + u^2)]\} ,
\end{eqnarray}

\begin{eqnarray}
\lefteqn{\frac{d\sigma}{dt}\left(d\overline{u}\to Q\overline{Q}
\left[{}^3\!S_1^{(8)}\right]W^-\right)
=\frac{d\sigma}{dt}\left(u\overline{d}\to Q\overline{Q}
\left[{}^3\!S_1^{(8)}\right]W^+\right)
=\frac{\pi \alpha_s^2 g^{\prime2}|V_{ud}|^2}{27 M^3 s^2 t^2 u^2}}
\nonumber\\
&&{} \times \{ m_W^4 (t^2 + u^2) - m_W^2 (s + t + u) (t^2 + u^2)
+ t u [2 s^2 + 2 s (t + u) + t^2 + u^2]\}. \qquad
\end{eqnarray}

\end{appendix}

\newpage
\begin{figure}[ht]
\begin{center}
\epsfig{figure=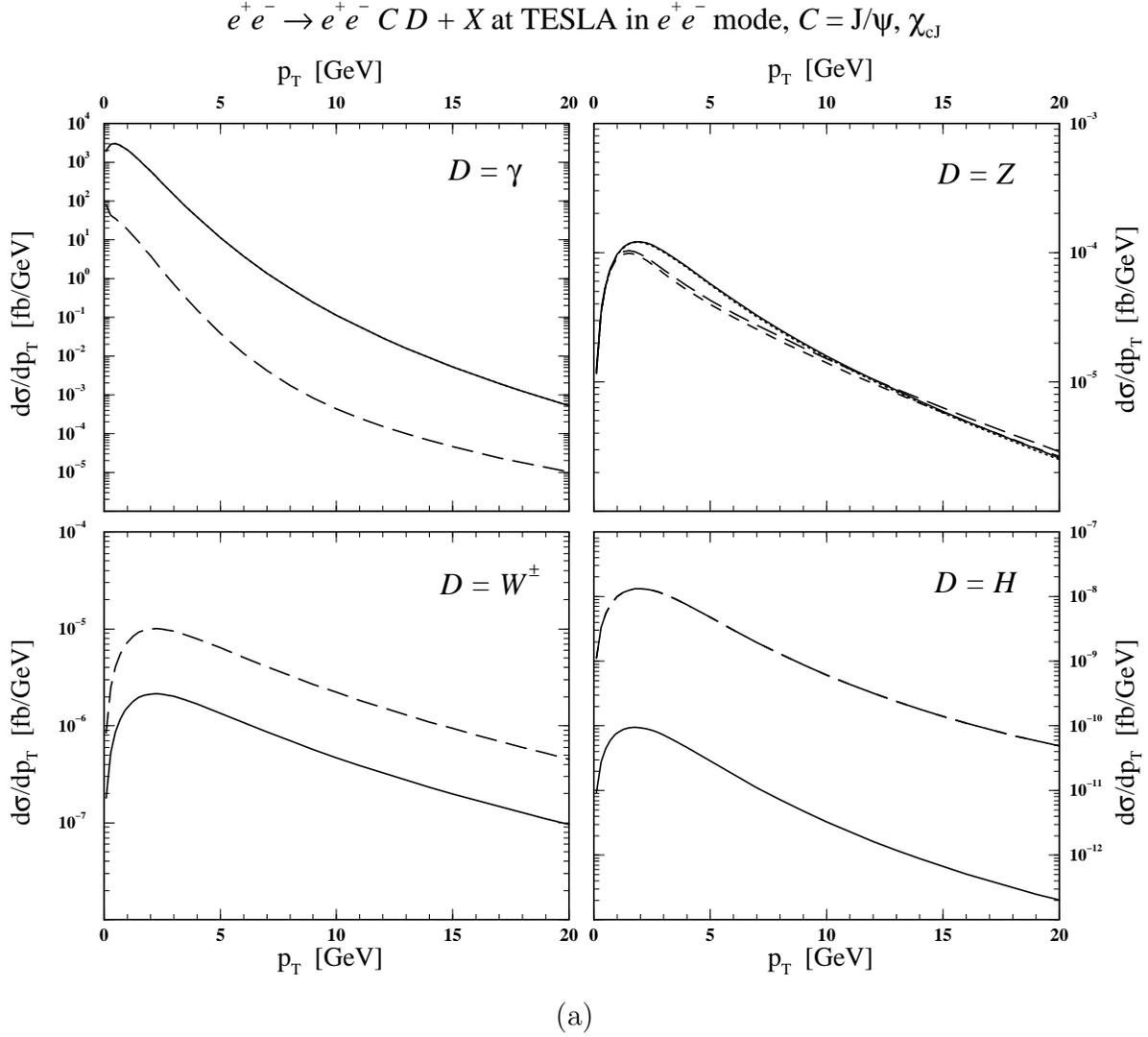,width=\textwidth}
(a)
\caption{(a) $p_T$ distributions $d\sigma/dp_T$ (in fb/GeV) and (b) $y_C$
distributions $d\sigma/dy_C$ (in fb) of $e^+e^-\to e^+e^-CD+X$, where
$C=J/\psi,\chi_{cJ}$ and $D=\gamma,Z,W,H$, at TESLA in the $e^+e^-$ mode.
It is summed over $C=\chi_{c0},\chi_{c1},\chi_{c2}$ and $D=W^+,W^-$.
In each figure, the CSM (dotted lines) and NRQCD (solid lines) predictions for
$C=J/\psi$ and the CSM (short-dashed lines) and NRQCD (medium-dashed lines)
ones for $C=\chi_{cJ}$ are shown separately.
Superposed short-dashed and medium-dashed lines appear as long-dashed lines.
\label{fig:ee}}
\end{center}
\end{figure}

\newpage
\begin{figure}[ht]
\begin{center}
\epsfig{figure=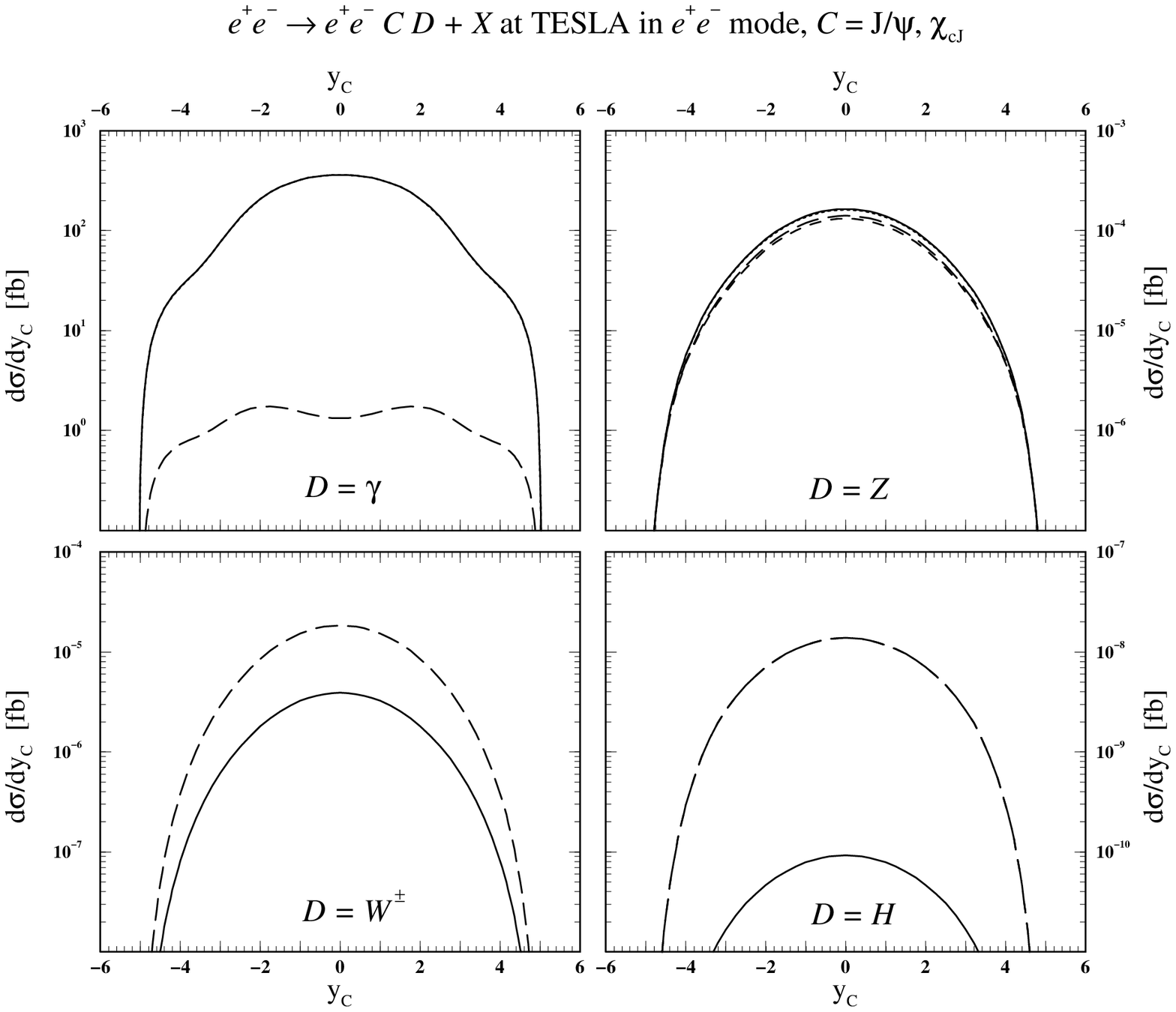,width=\textwidth}
(b)\\
Fig.~\ref{fig:ee} (continued).
\end{center}
\end{figure}

\newpage
\begin{figure}[ht]
\begin{center}
\epsfig{figure=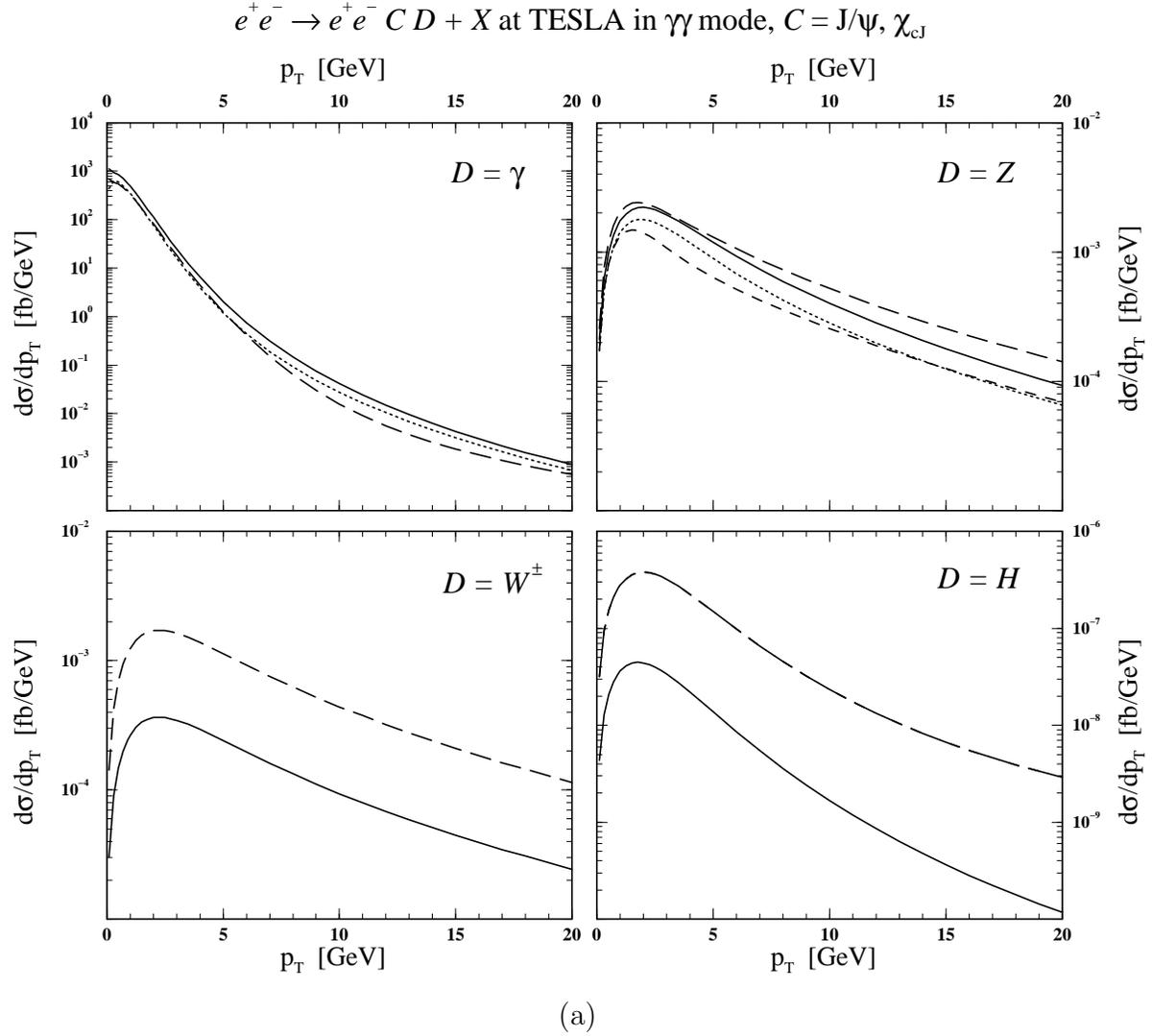,width=\textwidth}
(a)
\caption{Same as in Figs.~\ref{fig:ee}(a) and (b), but for
$e^+e^-\to e^+e^-CD+X$ at TESLA in the $\gamma\gamma$ mode.
\label{fig:gg}}
\end{center}
\end{figure}

\newpage
\begin{figure}[ht]
\begin{center}
\epsfig{figure=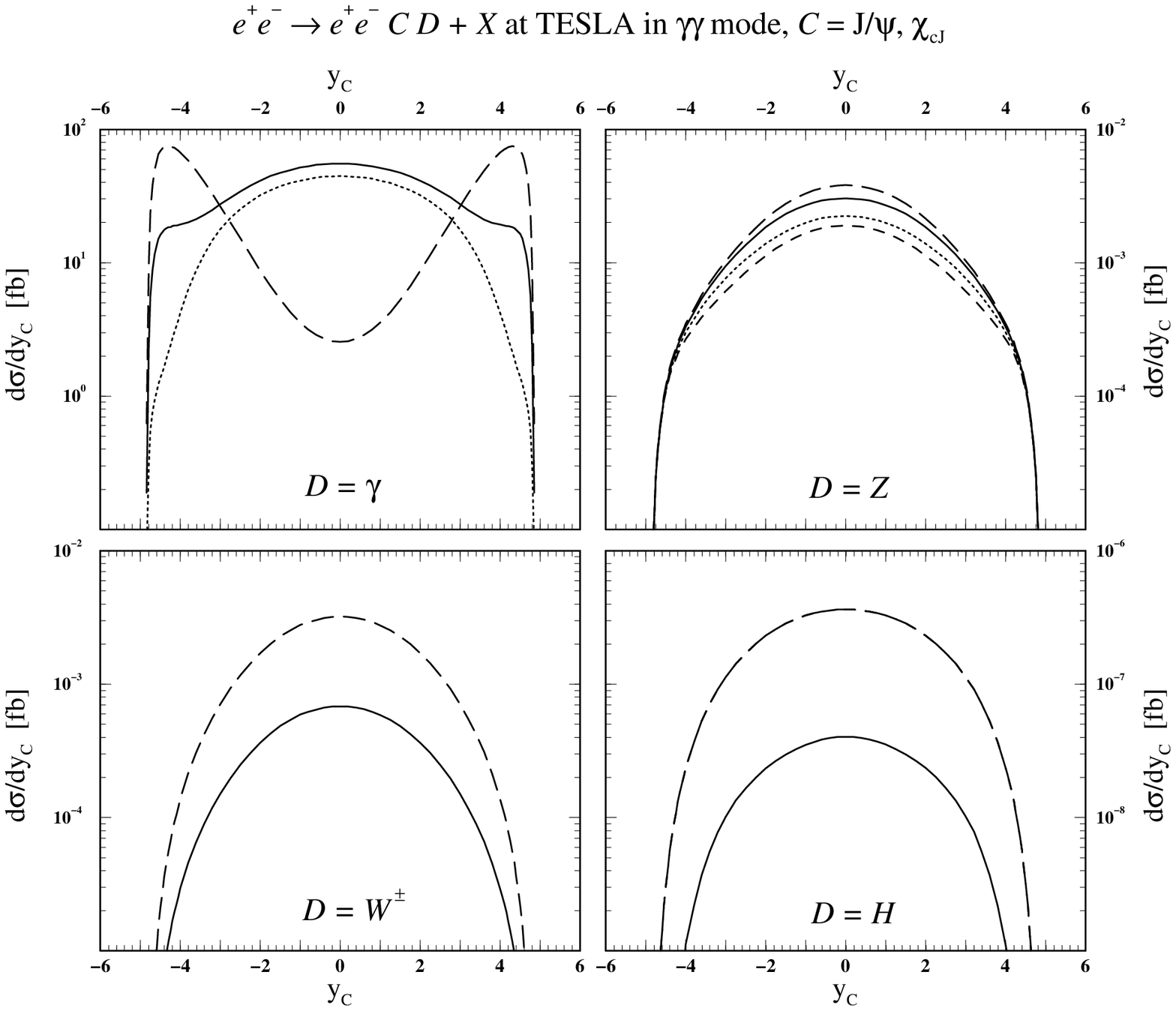,width=\textwidth}
(b)\\
Fig.~\ref{fig:gg} (continued).
\end{center}
\end{figure}

\newpage
\begin{figure}[ht]
\begin{center}
\epsfig{figure=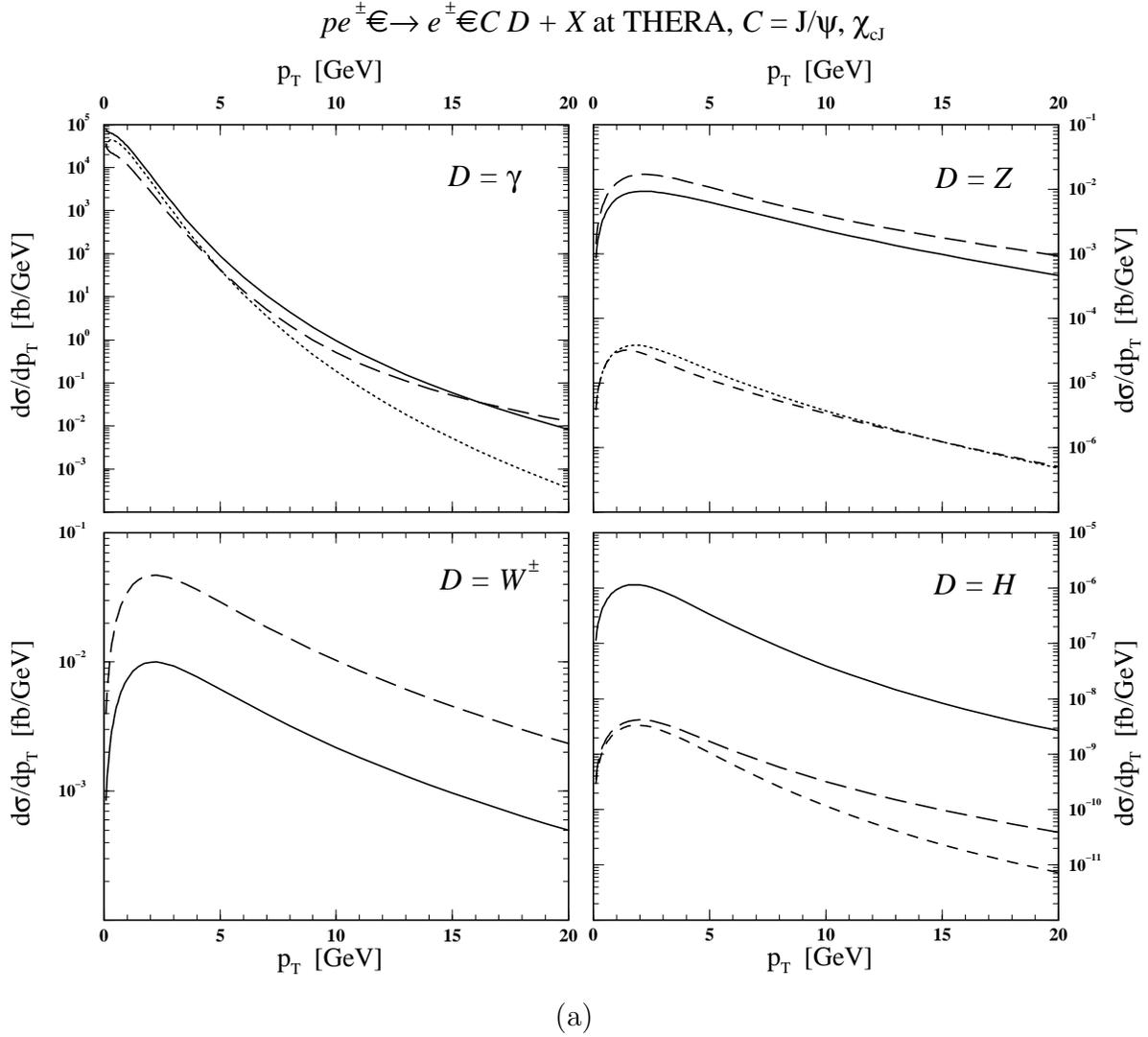,width=\textwidth}
(a)
\caption{Same as in Figs.~\ref{fig:ee}(a) and (b), but for
$pe^\pm\to e^\pm CD+X$ at THERA in the laboratory frame.
\label{fig:pe}}
\end{center}
\end{figure}

\newpage
\begin{figure}[ht]
\begin{center}
\epsfig{figure=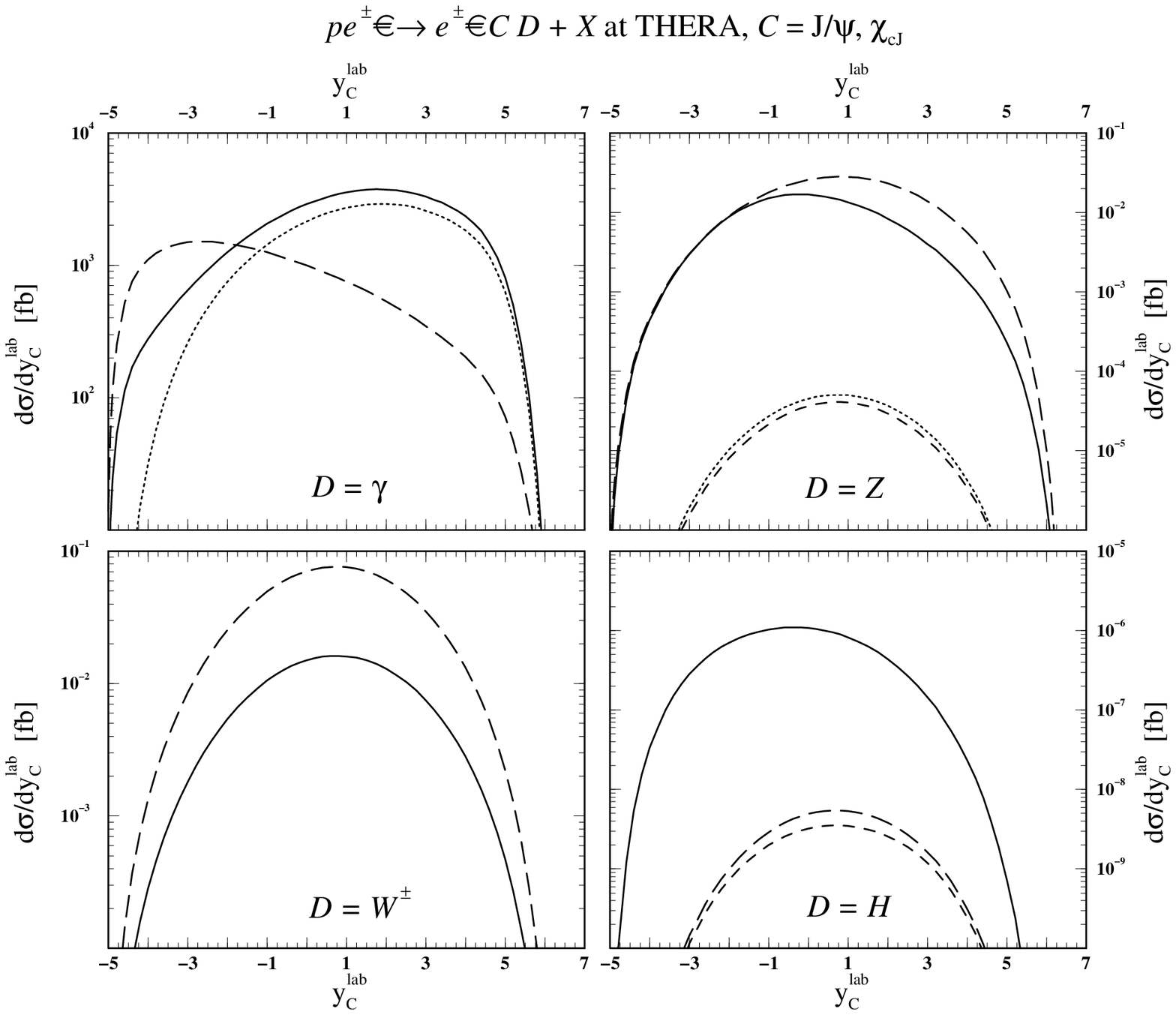,width=\textwidth}
(b)\\
Fig.~\ref{fig:pe} (continued).
\end{center}
\end{figure}

\newpage
\begin{figure}[ht]
\begin{center}
\epsfig{figure=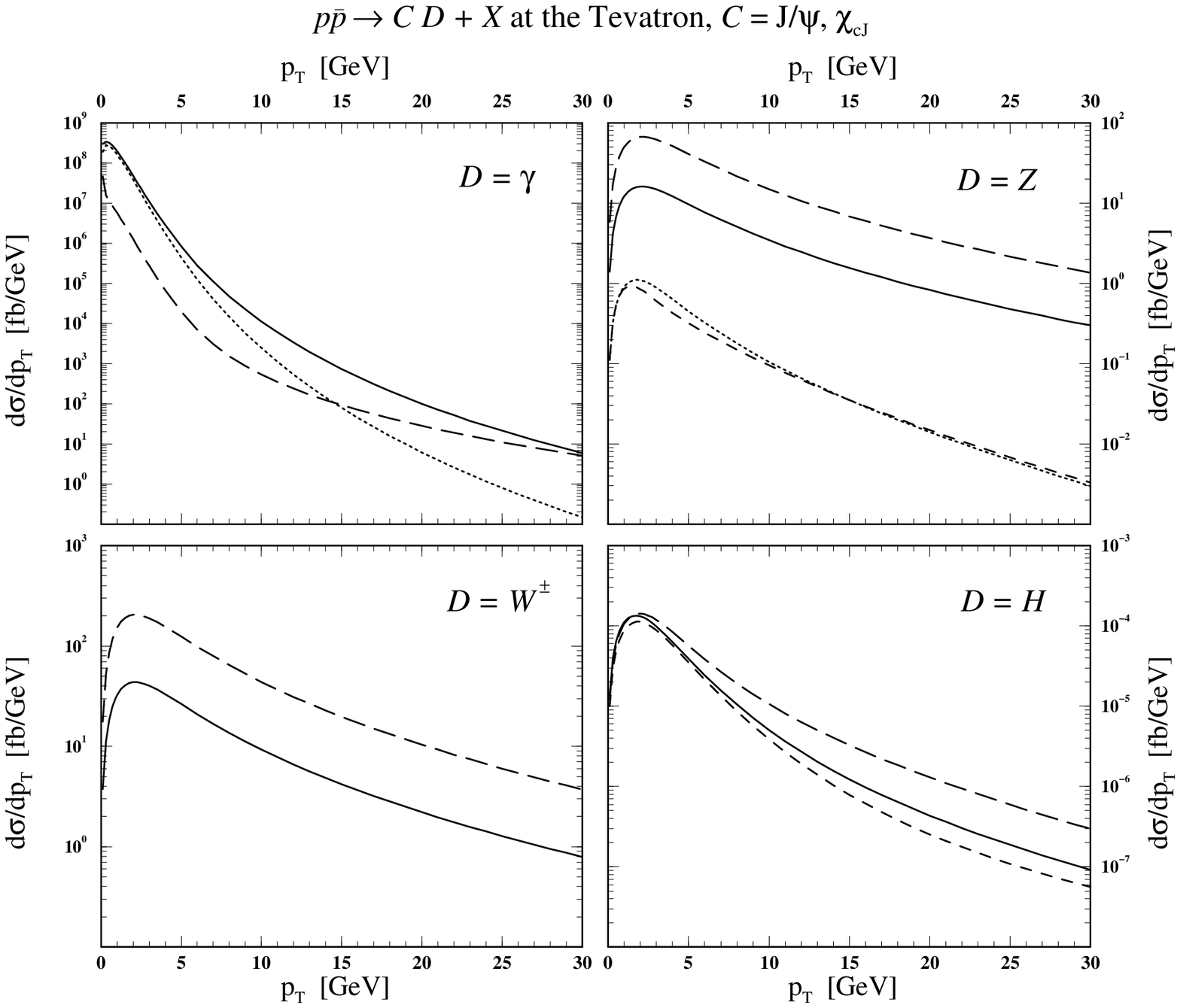,width=\textwidth}
(a)
\caption{Same as in Figs.~\ref{fig:ee}(a) and (b), but for
$p\overline{p}\to CD+X$ at the Tevatron.
\label{fig:pa}}
\end{center}
\end{figure}

\newpage
\begin{figure}[ht]
\begin{center}
\epsfig{figure=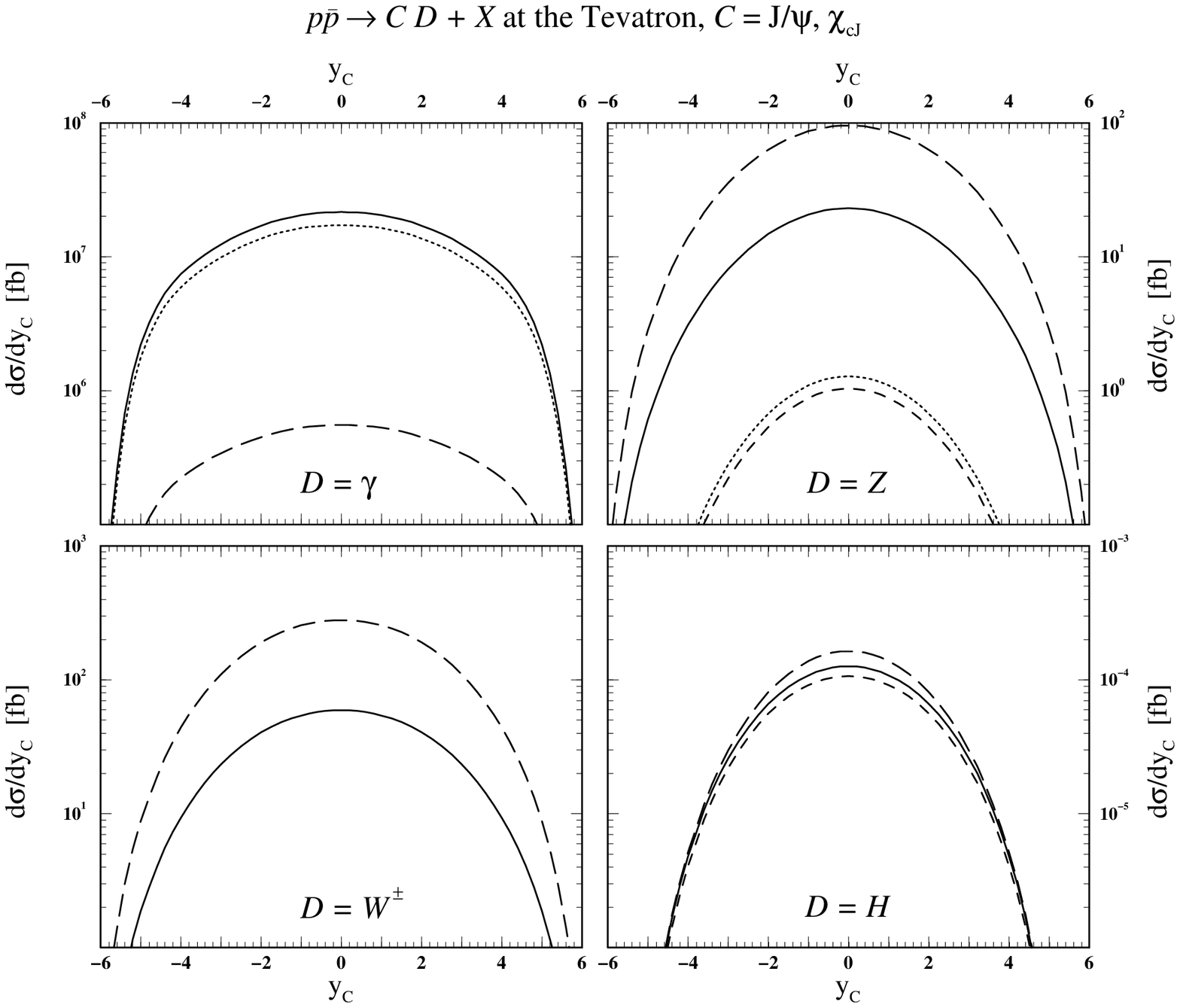,width=\textwidth}
(b)\\
Fig.~\ref{fig:pa} (continued).
\end{center}
\end{figure}

\newpage
\begin{figure}[ht]
\begin{center}
\epsfig{figure=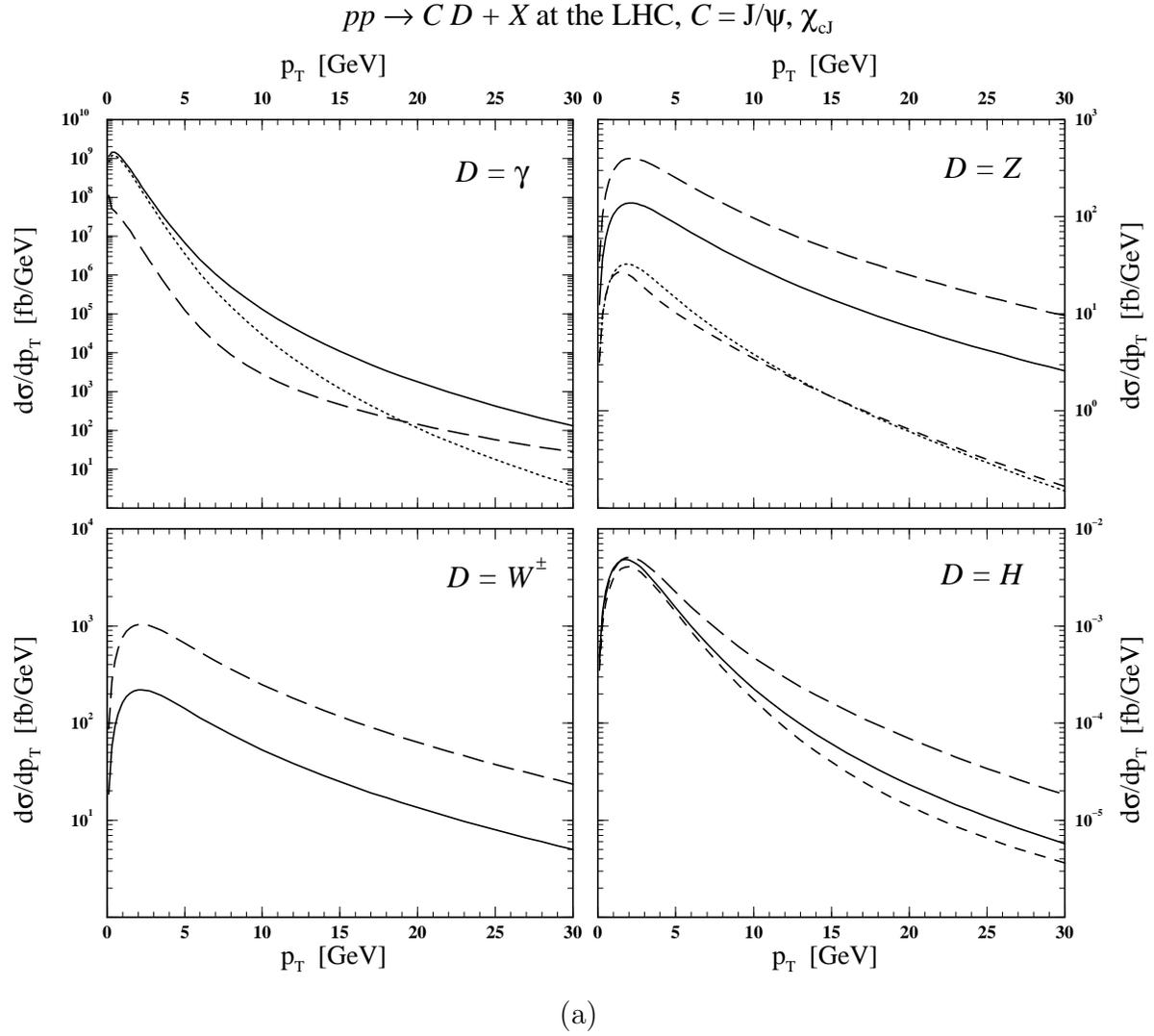,width=\textwidth}
(a)
\caption{Same as in Figs.~\ref{fig:ee}(a) and (b), but for $pp\to CD+X$ at the
LHC.
\label{fig:pp}}
\end{center}
\end{figure}

\newpage
\begin{figure}[ht]
\begin{center}
\epsfig{figure=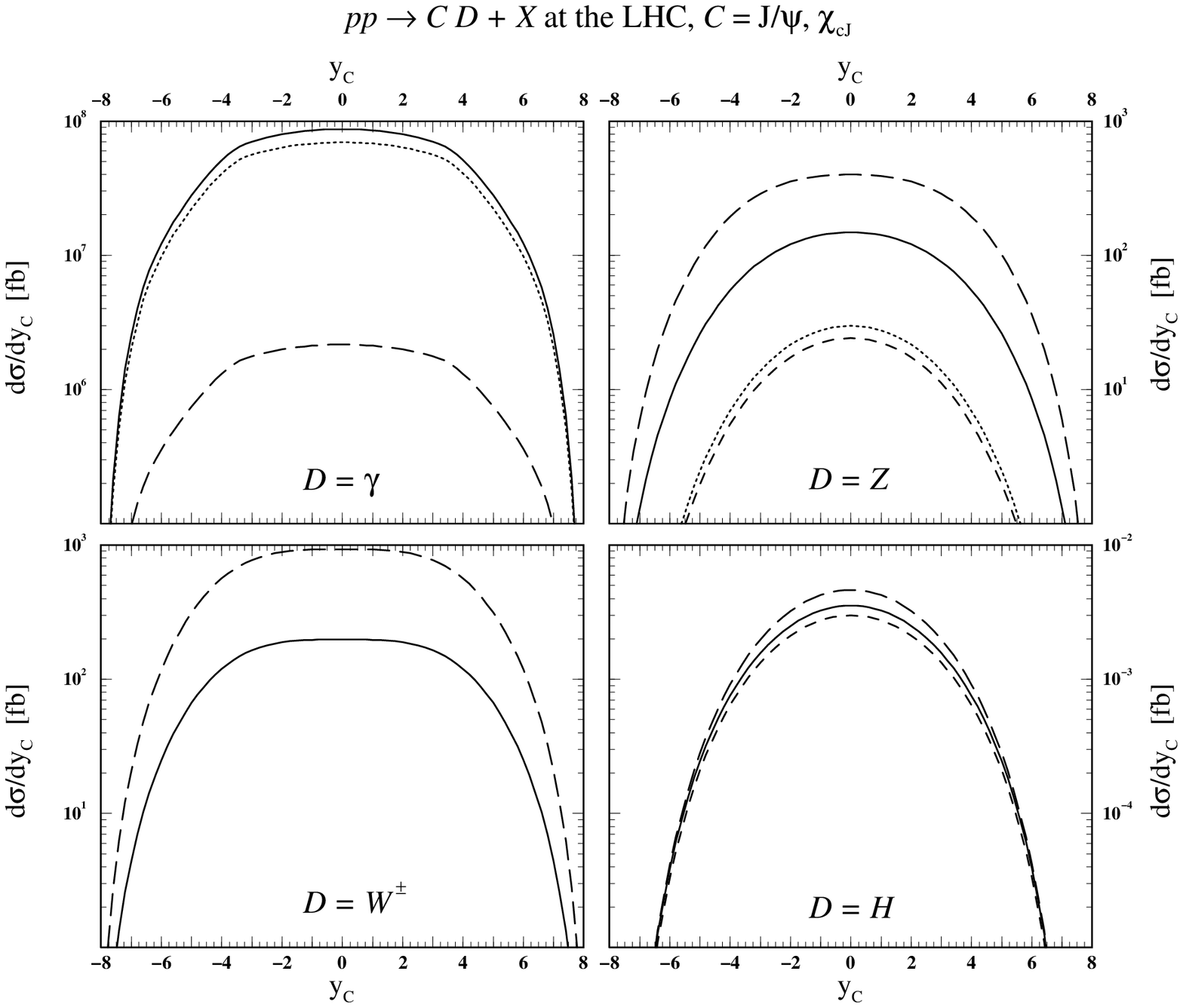,width=\textwidth}
(b)\\
Fig.~\ref{fig:pp} (continued).
\end{center}
\end{figure}

\end{document}